\pdfoutput=1
\documentclass[10pt, conference, letterpaper]{IEEEtran}
\usepackage[T1]{fontenc}
\usepackage[utf8]{inputenc}
\usepackage{amsmath}
\usepackage{xspace}
\usepackage{amsthm}
\usepackage{amssymb}
\usepackage{todonotes}
\usepackage{tabularx}
\usepackage[binary-units]{siunitx}
\usepackage[USenglish]{babel}
\usepackage[noend]{algpseudocode}
\usepackage{algorithmicx}
\usepackage{pgfplots}
\makeatletter
\let\MYcaption\@makecaption
\makeatother

\usepackage[font=footnotesize]{subcaption}

\makeatletter
\let\@makecaption\MYcaption
\makeatother
\usepackage{algorithm}
\usepackage{tikz}
\usetikzlibrary{arrows,automata, shapes, plotmarks, decorations.pathreplacing}
\usetikzlibrary{decorations.pathreplacing}
\usetikzlibrary{positioning, fit}

\usepackage{hyperref}
\usepackage{cleveref}

\usepackage[%
 backend=biber,
 style=ieee,
 url=false,
 isbn=false,
 doi=false,
 maxcitenames=1,
 mincitenames=1,
 maxbibnames=5,
]{biblatex}

\AtEveryBibitem{\clearfield{note}}
\AtEveryBibitem{\clearlist{location}}
\AtEveryBibitem{\clearfield{pages}}
\AtEveryBibitem{\clearfield{series}}
\AtEveryBibitem{\clearfield{month}}

\newcommand*{\eg}{e.g.\@\xspace}
\newcommand*{\ie}{i.e.\@\xspace}

\newcommand{\N}{\mathbb{N}}
\newcommand{\Pb}{\mathbb{P}}
\newcommand{\Ex}{\mathbb{E}}

\begin{document}
\title{Eclipsing Ethereum Peers with False Friends}

\author{
\IEEEauthorblockN{Sebastian Henningsen \quad\qquad Daniel Teunis \quad\qquad Martin Florian \quad\qquad Björn Scheuermann}
\IEEEauthorblockA{~\\
Weizenbaum-Institute for the Networked Society\\
Humboldt-Universität zu Berlin\\Berlin, Germany\\
}}
\maketitle
\begin{abstract}
Ethereum is a decentralized Blockchain system that supports the execution of Turing-complete smart contracts.
Although the security of the Ethereum ecosystem has been studied in the past, the network layer has been mostly neglected.
We show that Go Ethereum (Geth), the most widely used Ethereum implementation, is vulnerable to eclipse attacks, effectively circumventing recently introduced security enhancements.
\footnote{We responsibly disclosed the vulnerability to core Ethereum developers, see Sec.~\ref{sec:countermeasures} for a discussion of implemented countermeasures in Geth v1.9.0}.
Our false friends attack exploits the Kademlia-inspired peer discovery logic used by Geth and enables a low-resource eclipsing of long-running, remote victim nodes.
An adversary only needs two hosts in distinct /24 subnets to launch the eclipse, which can then be leveraged to filter the victim's view of the Blockchain.
We discuss fundamental properties of Geth's node discovery logic that enable the false friends attack, as well as proposed and implemented countermeasures.
\end{abstract}

\section{Introduction}
\label{sec:intro}

A dependable and secure network layer is vital for blockchain systems, since they build on the assumption of equal information at every peer~\cite{DBLP:conf/uss/HeilmanKZG15,DBLP:journals/comsur/TschorschS16}.
This assumption is violated if \emph{eclipse attacks} are possible.
In an eclipse attack, an adversary monopolizes the connections of a victim, effectively filtering the victim's view of the blockchain.
Eclipse attacks enable a variety of follow-up attacks such as double spending and stubborn mining~\cite{DBLP:conf/eurosp/NayakKMS16}.

Despite its important role, the network layer has so far received surprisingly little attention in systems like Bitcoin~\cite{nakamoto2008bitcoin,DBLP:journals/comsur/TschorschS16} or Ethereum~\cite{wood2014ethereum}, leading to vulnerabilities.

In \cite{DBLP:journals/iacr/MarcusHG18}, a low-cost eclipse attack has been proposed that exploits the Kademlia-inspired \cite{DBLP:conf/iptps/MaymounkovM02} peer discovery logic of \emph{Go Ethereum (Geth)}, the official reference implementation of Ethereum\footnote{
  Other clients might also be vulnerable. We focus on Geth as it is estimated to be used in roughly \SI{76}{\percent} of clients~\cite{DBLP:conf/imc/KimMMMMB18}.
}.
The attack is mounted after a victim node has been restarted and is based on flooding the node's discovery table with Sybil~\cite{douceur2002sybil} nodes.
Generating a new node ID, and henceforth a new Sybil node, involves only an ECDSA key pair generation, which makes the attack lightweight.
As an answer to the discovered attack vector, Geth $\geq$ v1.8.0 introduces several countermeasures to increase the difficulty and necessary resources to flood the complete discovery table.

However, as we show in this paper, eclipse attacks on Geth are still possible with very limited effort.
We propose the \emph{false friends} attack that enables the eclipsing of current Ethereum nodes.
Despite the subnet restrictions implemented in Geth v1.8.0, we only need two IP addresses from distinct /24 subnets for a successful attack.
Additionally, and in contrast to \cite{DBLP:journals/iacr/MarcusHG18}, we do not necessarily require a restart of the victim node since peer churn is high and existing connections will eventually be dropped.
Instead of overwriting the complete discovery table with Sybil nodes, we subtly insert adversarial nodes with carefully selected node IDs,
exploiting the interplay between peer discovery and connection management.

Geth chooses new peers either by directly selecting nodes from its discovery table or by starting a Kademlia-style lookup to a random target, which yields new node contact information.
We compromise both mechanisms, in slightly different ways.
We insert a limited number of Sybil nodes into the victim's discovery table, with an activity pattern that favors these Sybils when new connections are set up.
For new contacts resulting from lookup operations, we pre-compute a large number of node IDs and present tailored choices when queried during a lookup, effectively offering ``better'' (albeit false) peers than all honest nodes visible to the victim.

Existing connections have to be terminated before the available slots can be filled with adversarial nodes.
However, as we noticed through measurements, peer connections are terminated regularly without additional intervention: \SI{95}{\percent} of connections longer than \SI{60}{\second} were shorter than \SI{5.5}{\day}
We were, in effect, able to successfully eclipse a live node by actively waiting and incrementally injecting adversary nodes into the victim's peer lists\footnote{
  Without attacking nodes operated by other network participants, and without risking any harm for the Ethereum network, of course.
}.
Our measurements indicate that a targeted false friends attack on a specific Ethereum node can be successfully completed within a few days.

In summary, the contributions of our paper are:
\begin{itemize}
  \item The discovery, description and evaluation of the \emph{false friends} attack, an eclipse attack on current Geth versions that exploits fundamental properties of Geth's peer discovery logic
  \item A description and theoretical analysis of Ethereum's network layer management algorithms, based on an analysis of the Geth codebase; previously available information is scarce.
  \item The discussion of possible and implemented countermeasures---both easy fixes to prevent the presented attack and ideas for tackling the fundamental challenge of securing Ethereum's overlay network.
\end{itemize}

The rest of this paper is structured as follows.
\Cref{sec:networkLayer} describes the high-level network architecture of Geth and \Cref{sec:dht} gives a detailed explanation of the Kademlia-inspired peer discovery.
In \Cref{sec:eclipseAttack} we then present the actual false friends attack, \ie how we exploit Geth's peer discovery logic.
We present our analytical evaluation and measurements in Sections~\ref{sec:analysis} and~\ref{sec:eval}.
\Cref{sec:countermeasures} discusses possible countermeasures against the presented attack.
We conclude the paper with a summary of related work (\Cref{sec:rw}) and concluding remarks (\Cref{sec:conclusion}).

\section{Background: the Ethereum Network Stack}
\label{sec:networkLayer}

In the following, we introduce the overall architecture of Ethereum's Peer-to-Peer network as implemented in Go Ethereum.
Unlike similar descriptions in related works~\cite{DBLP:conf/imc/KimMMMMB18,DBLP:journals/iacr/MarcusHG18}, the information presented here uses a naming of high-level components that is more strongly aligned with the official Ethereum terminology.

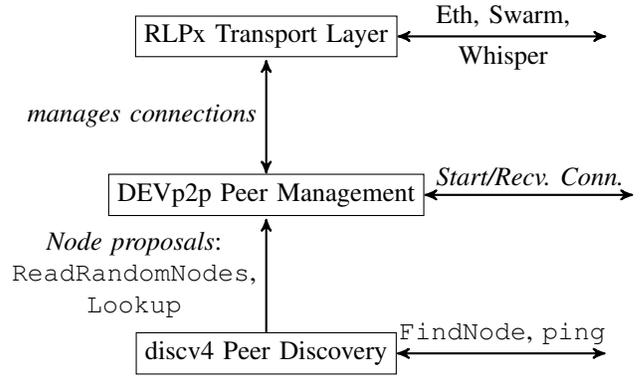
\begin{figure}
\centering
\begin{tikzpicture}[>=stealth',shorten >=1pt,auto, node distance = 6em, minimum width=3em]
\def\hDist{8em}

  \node[draw, rectangle, color=black] (pd) {discv4 Peer Discovery};
  \node[draw, rectangle, color=black] [above of=pd] (p2p) {DEVp2p Peer Management};
  \node[draw, rectangle, color=black] [above of=p2p] (sub) {RLPx Transport Layer};

  \coordinate[right= \hDist of pd] (pdr);
  \coordinate[right= \hDist of p2p] (p2pr);
  \coordinate[right= \hDist of sub] (subr);

  \path [->] (pd) edge [draw, thick, left] node [align=center] {\textit{Node proposals}:\\ \texttt{ReadRandomNodes},\\ \texttt{Lookup}} (p2p);
  \path [<->] (p2p) edge [draw, thick, left] node {\textit{manages connections}} (sub);
  
  \path [<->] (pd) edge [draw, thick, above] node {\texttt{FindNode}, \texttt{ping}} (pdr);
  \path [<->] (p2p) edge [draw, thick, above] node {\textit{Start/Recv. Conn.}} (p2pr);
  \path [<->] (sub) edge [draw, thick, above] node[midway,above] {Eth, Swarm,} node[midway,below] {Whisper} (subr);
\end{tikzpicture} 
\caption{Overview of the Ethereum Network Stack.}
\label{fig:networkStack}
\end{figure}

The overall network architecture of Ethereum is summarized in \Cref{fig:networkStack}.
Ethereum's network layer consists of four major components, namely: \emph{discv4} for node discovery; \emph{RLPx} as a secure transport layer; \emph{DEVp2p} for session management on top of RLPx and the actual \emph{Ethereum protocol} (\emph{eth}) which runs on top of DEVp2p.
The Whisper protocol (for decentralized applications) and the \emph{Swarm} protocol (for decentralized file storage) are other subprotocols on top of DEVp2p.

DEVp2p not only provides the foundation for the Ethereum protocol and other application protocols,
it also manages connections to other peers which form the overlay on which blocks and transactions are distributed.
Geth by default has a total of 25 TCP connections to other peers speaking the Ethereum protocol.
Of these 25 slots, 17 are reserved for inbound connections (initiated by other peers), whereas the remaining 8 are allocated for outbound connections.
In this case, inbound means that a remote peer sent a \texttt{SYN}-packet to start a TCP connection with the local peer.
No further restrictions apply to inbound connections; if an inbound slot is available Geth simply accepts any connecting peer that supports the Ethereum protocol and operates on the same network (main, testing, etc.).
The 8 outbound slots are therefore especially important, as they are the most difficult ones to get under control for an attacker mounting an eclipse attack.

In contrast to DEVp2p, the discv4 node discovery stores information about \emph{all} node types in the overlay.
This includes nodes without support for the Ethereum protocol (which is a perfectly valid case in the design logic of Ethereum's protocol stack).
The discv4 node discovery is inspired by the Kademlia DHT~\cite{DBLP:conf/iptps/MaymounkovM02}.
Information about known overlay nodes is stored in a table separated into so-called $k$-buckets (or simply \emph{buckets}, in the following).

\begin{figure}
  \centering
\begin{tikzpicture}[node distance = 2cm, auto]
      \tikzstyle{decision} = [diamond, draw, fill=blue!20,
      text width=4.5em, text badly centered, node distance=3cm, inner sep=0pt]
      \tikzstyle{block} = [rectangle, draw, fill=blue!20, 
      text width=5em, text centered, rounded corners, minimum height=4em]
      \tikzstyle{line} = [draw, -latex']
  \node[block] (init) {$n > 0$ free outbound slots};
  \node[block, below of=init] (RRN) {Read $\lfloor n/2 \rfloor$ nodes from table};
  \node[decision, below of=RRN] (lookupempty) {Lookup-buffer empty?};
  \node[block, right=1cm of lookupempty] (readlookup) {Read $n - \lfloor n/2 \rfloor$ nodes from lookup-buffer};
  \node[block, left=1cm of lookupempty] (dolookup) {Lookup to random target};

  \path[line] (init) -- (RRN);
  \path[line] (RRN) -- (lookupempty);
  \path[line] (lookupempty) -- node[near start] {yes} (dolookup);
  \path[line] (lookupempty) -- node[near start] {no} (readlookup);
\end{tikzpicture} 

  \caption{How outbound connections are established.}
  \label{fig:outboundFlowChart}
\end{figure}
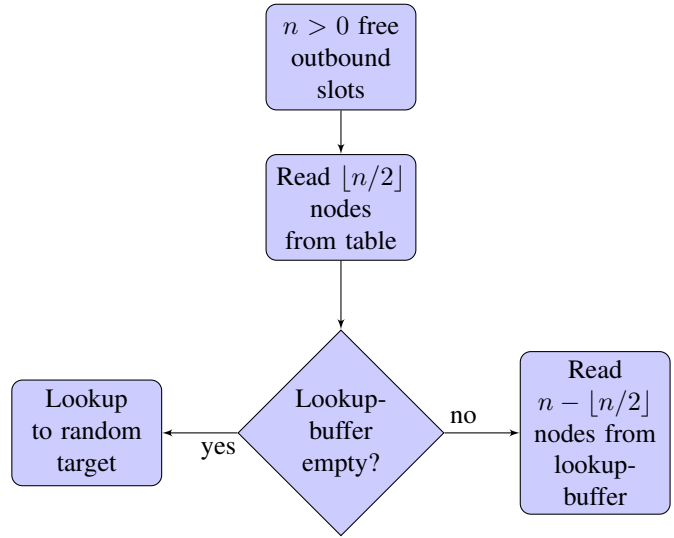

The outbound connections are established to nodes that are returned from the discovery table.
Every time not all outbound slots are occupied, the DEVp2p peer management requests the discovery table in two distinct fashions depicted in \Cref{fig:outboundFlowChart}.

First, half of the \emph{currently empty} slots (rounded down) are filled with a direct request to the discovery table via the function \texttt{ReadRandomNodes}.
Second, the remaining slots are filled from the lookup-buffer, which holds the result of a Kademlia-like lookup to a random target ID.
Note that this procedure is repeated \emph{every time an outbound slot becomes available}.
Therefore, Geth fills half of the \emph{currently available} slots with each mechanism.
Depending on the situation, this skews the distribution of outbound connections towards either mechanism.
If only one slot becomes available at a time, the lookup-buffer is favored (\texttt{ReadRandomNodes} gets $\lfloor0.5\rfloor = 0$ slots).
Otherwise, if two lookup-buffer slots become available repeatedly, \texttt{ReadRandomNodes} is favored in comparison to the lookup-bufer.

Our false friends attack exploits these two interfaces between node discovery and peer management.
The discovery table therefore constitutes a particularly important component of Ethereum for our purposes, and deserves a closer look.

\section{Node Discovery and Selection}
\label{sec:dht}

Ethereum's node discovery table largely resembles a \emph{Kademlia}~\cite{DBLP:conf/iptps/MaymounkovM02} routing table.
Unlike in Kademlia, however,
its sole purpose is to manage a set of known nodes which serves as a basis for establishing connections in DEVp2p.
In \cite{DBLP:conf/icdcn/LocherMSW10} it is conjectured that Kademlia was chosen as a basis due to future plans to shard the blockchain, \ie, to partition it over the network.

\subsection{Kademlia in a Nutshell}

Kademlia is a UDP-based, peer-to-peer distributed hash table (DHT) that is used to locate decentrally stored data efficiently~\cite{DBLP:conf/iptps/MaymounkovM02}.
Each node is uniquely identified by its randomly generated 160-bit node ID.
A data item stored in the DHT is found by its key, which is simply a 160-bit hash of the data itself.
Hence, node IDs and keys share the same representation; in the following we use the term ID for both.
Kademlia leverages this by storing data at nodes whose node ID is ``close'' to the data's key.
Closeness is defined as the bitwise XOR between two IDs, taken as an integer value, \ie, $d(x, y) = x \oplus y$.
A node stores its known neighbors in so-called \emph{$k$-buckets} which partition the known network based on the local node's ID.
Every $k$-bucket (or simply bucket) stores up to $k$ neighbors.
Intuitively, node IDs are treated as the leaves of a binary tree and each bucket stores a distinct branch of the tree.
Bucket $i$ stores nodes whose distance is in $[2^i, 2^{i+1})$, which effectively corresponds to the length of the common prefix between two node IDs.

Items in the DHT and nodes on the network are located by so-called \emph{lookups}.
A lookup successively queries nodes that are closest to the desired target ID (key or node ID).

\subsection{Node IDs}
\label{subsec:nodeIDs}

As in Kademlia, node IDs in Ethereum serve as public identifiers for each node in the Ethereum network.
A node ID in Ethereum is a marshaled 512-bit ECDSA public key.
However, distance computations only operate on Keccak256-hashes~\cite{bertoni2013keccak} of node IDs, effectively yielding node IDs with 256-bit length.
When referring to node IDs in the following we implicitly mean hashed ECDSA public keys.
Node IDs are supposed to be static, as stated in the official Ethereum documentation:
\begin{quote}\it
  Each node is expected to maintain a static private key which is saved and restored between sessions. It is recommended that the private key can only be reset manually [...].\footnote{\href{https://github.com/ethereum/devp2p/blob/6504d410bc4b8dda2b43941e1cb48c804b90cf22/rlpx.md}{https://github.com/ethereum/devp2p/blob/\linebreak6504d410bc4b8dda2b43941e1cb48c804b90cf22/rlpx.md}, accessed 15.04.19}
\end{quote}
It is easy to generate and use many different identities by creating ECDSA key pairs.

\subsection{Buckets and Log-Distance-Metric}
\label{subsec:tableBucketsDist}

The buckets of Geth's discovery table hold up to $k = 16$ nodes each.
Exactly as in Kademlia, the nodes in each bucket share a common property:
the distance, according to some metric, between their node ID and the local node's ID is the same.
\cite{DBLP:conf/imc/KimMMMMB18,DBLP:journals/iacr/MarcusHG18} state that Ethereum uses the so-called \emph{log-distance} metric.
We argue that this metric is identical to the distance metric used to determine buckets in Kademlia.
The log-distance between two hashes can be defined as $\lfloor \log_2 (\bar{N_1} \oplus \bar{N_2})\rfloor$, or equivalently, 255 - the length of their common prefix -- which is exactly how buckets are organized in Kademlia.
Due to the uniqueness assumption of node IDs, this yields $| \{0, ..., 255\} | = 256$ possible distances.

In response to the eclipse attack by~\cite{DBLP:journals/iacr/MarcusHG18}, Geth $\geq$ 1.8.0 restricts the number of buckets to $17$, starting from the furthest distance of 255 to the minimum possible log-distance of $239$.

The log-distance metric leads to a skewed distribution of nodes between buckets: most of the lower buckets are empty, since the probability to fall into a specific bucket decays exponentially with the associated distance~\cite{DBLP:conf/iptps/MaymounkovM02}.

\subsection{Entering a Bucket}

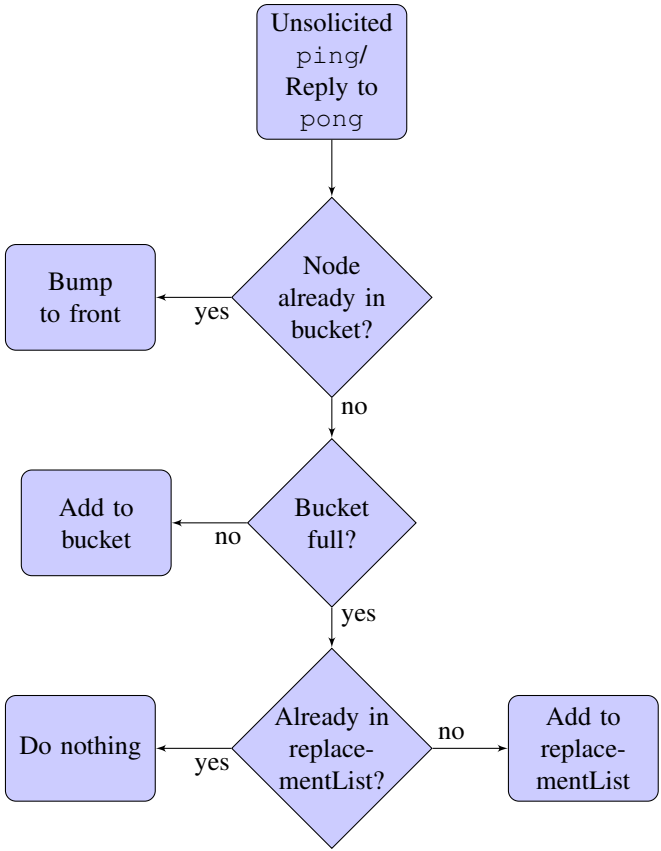
\begin{figure}
\begin{tikzpicture}[node distance = 2cm, auto]
      \tikzstyle{decision} = [diamond, draw, fill=blue!20,
      text width=4.5em, text badly centered, node distance=3cm, inner sep=0pt]
      \tikzstyle{block} = [rectangle, draw, fill=blue!20, 
      text width=5em, text centered, rounded corners, minimum height=4em]
      \tikzstyle{line} = [draw, -latex']
  \node[block] (init) {Unsolicited \texttt{ping}/\\Reply to \texttt{pong}};
  \node[decision, below of=init] (inBucket) {Node already in bucket?};
  \node[block, left=1cm of inBucket] (bump) {Bump to front};
  \node[decision, below of=inBucket] (bucketFull) {Bucket full?};
  \node[block, left=1cm of bucketFull] (addBucket) {Add to bucket};
  \node[decision, below of=bucketFull] (inReplacement) {Already in replacementList?};
  \node[block, left=1cm of inReplacement] (doNothing) {Do nothing};
  \node[block, right=1cm of inReplacement] (addReplacement) {Add to replacementList};
  \path[line] (init) -- (inBucket);
  \path[line] (inBucket) -- node[near start] {yes} (bump);
  \path[line] (inBucket) -- node[near start] {no} (bucketFull);
  \path[line] (bucketFull) -- node[near start] {no} (addBucket);
  \path[line] (bucketFull) -- node[near start] {yes} (inReplacement);
  \path[line] (inReplacement) -- node[near start] {yes} (doNothing);
  \path[line] (inReplacement) -- node[near start] {no} (addReplacement);

\end{tikzpicture} 

  \caption{How nodes enter buckets.}
  \label{fig:bucketflowChart}
\end{figure}

A local node learns of neighboring nodes either by receiving an unsolicited \texttt{ping} packet or in the course of a lookup operation.
The lookup process also initiates a \texttt{ping}/\texttt{pong} exchange, which then triggers the node to be added to the discovery table.
In any case, before a node enters the discovery table, a number of checks are performed which are depicted in \Cref{fig:bucketflowChart}.
Assume that the local node receives either a \texttt{ping} packet or a \texttt{pong} reply to a previously sent \texttt{ping}.
Two cases are to be distinguished:
first, the node may already be in its respective bucket; in this case it is simply moved to the first position.
This induces a ``least recently active'' sorting of the nodes within a bucket~\cite{DBLP:conf/iptps/MaymounkovM02}, where activity simply means sending (responding to) a \texttt{ping}-packet.
Second, in case the node is not already in a bucket, it is added if the bucket is not full.
If the bucket is already full, candidate nodes are stored in a so-called \emph{replacement list} that stores up to ten nodes.

Every \SI{5}{\second} (on average), the last node of a random bucket is pinged and replaced with a random node from the respective replacement list if it fails to respond.
In contrast to buckets, the replacement list is a simple FIFO queue that evicts the last entry every time a previously unknown node is added to the list.
Last but not least, a node is only added to its respective bucket (or replacement list) if it meets certain IP address restrictions:
Geth restricts the number of IP addresses coming from the same /24 subnet to two per bucket, and to ten in the whole discovery table.
\subsection{FindNode-Requests}
\label{subsec:findnode}

Lookups in Ethereum are used to discover new peers.%
\footnote{FindNode requests are also used to populate the discovery table and to resolve node IDs to IP addresses~\cite{DBLP:journals/iacr/MarcusHG18}, which is outside the scope of this paper.}.
These lookups are performed iteratively by sending so-called \emph{FindNode requests}, to which the recipient answers with a \texttt{neighbors} packet containing information about nodes from its discovery table.

The most important use of lookups in our scenario is to populate the so-called \emph{lookup-buffer}.
As already outlined, the lookup-buffer is one of two methods by which the DEVp2p subsystem finds new nodes to connect to.
When the lookup-buffer is empty, Geth populates it by starting a lookup to a random target.
That is, it sends a FindNode request to peers that are ``close'' to the random target.
For lookups, Ethereum does \emph{not} use the log-distance metric for sorting the buckets (cf. \Cref{subsec:tableBucketsDist}) but instead uses the plain xor-metric.
To illustrate, let $\bar{N_1}, \bar{N_2}$ be two node IDs and $t$ a (random) target ID.
For Ethereum (and likewise Kademlia), $\bar{N_1}$ is closer to $t$ than $\bar{N_2}$ iff $\bar{N_1} \oplus t < \bar{N_2} \oplus t$, where $\oplus$ denotes the bitwise xor-operation and the result is taken as the binary representation of an unsigned integer.
To ease notation, we define the abbreviation $<_t$ as $\bar{N_1} <_t \bar{N_2} :\Leftrightarrow (\bar{N_1} \oplus t < \bar{N_2} \oplus t$).

The iterative lookup procedure to populate the lookup-buffer is visualized in \Cref{alg:lookupBuffer}.
First, a random target ID $t$ is chosen.
Subsequently, all known peers from the discovery table are sorted according to $<_t$,
effectively yielding the 16 peers that are closest to the random target $t$.
In a next step, a FindNode request is sent to each of these 16 peers, asking them for their respective neighbors that are closest to $t$.
If successful, each queried peer will answer with a \texttt{neighbors} packet containing up to 12 peers (1280 byte).
All received neighbors are combined, sorted by $<_t$, and again restricted to the 16 peers closest to the random target $t$. This yields the result set of the first round.
This process is iterated until the result set eventually stabilizes.
If a result set contains the same peers as the result set of the previous iteration, the procedure terminates.

\begin{algorithm}
  \caption{Populate Lookup-buffer}
  \label{alg:lookupBuffer}
  \begin{algorithmic}
    \State $t \gets \text{random node ID}$
    \State $N_0 \gets \{16 \text{ closest known peers to } t\}$
    \Loop
      \For{$o_i \in N_0$}
        \State $F_i \gets \{\text{closest peers of } o_i \text{ to } t\text{, as returned by } o_i \}$
      \EndFor
      \State $N \gets N_0 \cup \bigcup_{i=0}^{15} F_i$\\
      \State $N_1 \gets \text{sort}(N, t)[0:15]$ // 16 closest to $t$
      \If{$N_0 = N_1$}
        \Return $N_0$
      \Else
        \State $N_0 \gets N_1$
      \EndIf
    \EndLoop
  \end{algorithmic}
\end{algorithm}

\section{The False Friends Attack}
\label{sec:eclipseAttack}

After having established the necessary background in the previous sections, we now describe the details of our \emph{false friends} attack, with an in-depth analysis of the attack following in \Cref{sec:analysis}.

To eclipse a victim, its 8 slots for outbound connections as well as the 17 slots for inbound ones have to be filled with adversarial nodes.
The inbound connections slots can easily be filled since Geth does not impose any restrictions on inbound connections.
Hence, it suffices to start multiple Geth instances on different ports and configure them to repeatedly connect to the victim.
Due to the lack of restrictions, \emph{one} host with one IP address is enough to fill the inbound slots.
Note that these Geth instances do \emph{not} need to actively participate in block and transaction distribution.

To fill the outbound connection slots, we have to make sure that only adversarial nodes are proposed to the DEVp2p peer management via the two mechanisms (cf. \Cref{sec:networkLayer}).
Whereas~\cite{DBLP:journals/iacr/MarcusHG18} fills the whole discovery table with Sybil nodes to ensure that only these are proposed to the peer management, we leverage properties of the peer selection mechanism to achieve the same result with only one Sybil node per neighbor table bucket.
It suffices to have one Sybil node in each bucket of the neighbour table to make sure that only adversarial nodes are returned to the peer management.
This effectively circumvents the implemented countermeasures and still needs very little resources (two IP addresses in distinct /24 subnets).

In contrast to~\cite{DBLP:journals/iacr/MarcusHG18,DBLP:conf/uss/HeilmanKZG15} we do not necessarily require a restart of the victim node for our attack to be successful, though it speeds up the attack.
In both cases, an adversary has to wait until existing connections are terminated for other reasons, such as timeouts.
Our measurements (discussed in \Cref{sec:eval}) indicate that connections on the Ethereum network are rather short-lived and a successful attack against a non-restarting victim node is therefore possible in a matter of days.

Our attack is facilitated by the fact that countermeasures 2 and 3 from~\cite{DBLP:journals/iacr/MarcusHG18} were not implemented in Geth.
Countermeasure 2, a fixed mapping between the IP address and ECDSA key would raise the requirements for a false friend attack to 25 unique IP addresses (one for each connection slot).
While moderately increasing the necessary resources for an adversary, the impediment for the network is significant, since multiple Geth instances behind a NAT would be impossible.
Countermeasure 3, making the mapping of IDs to buckets in the neighbour table secret, is a theoretically viable mechanism to drastically raise the bar for an attacker.
However, one would lose the benefits of Kademlia in case routing becomes relevant; this might be the reason why the developers chose not to implement countermeasure 3.

\subsection{Taking Over \texttt{ReadRandomNodes}}
\label{subsec:readRandomNodes}

The function \texttt{ReadRandomNodes} returns (per default) at most 4 nodes from the discovery table, which are then used by the peer management to establish outbound connections.
Most importantly, \texttt{ReadRandomNodes} only returns the \emph{head} of randomly chosen buckets.
Presumably, this design choice is due to the implicit sorting by activity within a bucket (cf. \Cref{sec:dht}): peers in the front of a bucket are more active and/or have a better latency than the others and are therefore favorable to connect to.
This behavior can easily exploited by an adversary, since the sorting by activity merely requires the adversary to regularly send a \texttt{ping}-packet to stay ahead of the other peers.
Therefore, it is sufficient for an attacker to populate each bucket with \emph{one} node instead of the whole discovery table.
To this end, an adversary repeatedly generates new ECDSA key pairs, computes the node ID and checks whether this particular ID is mapped to the desired bucket.

Current versions of Geth maintain 17 buckets and implement an IP-based restrictions such that at most 2 nodes from the same /24 subnet can be included in the same bucket and at most 10 nodes from the same /24 subnet can be in the whole discovery table.
With these current properties of Geth, only \emph{two} IPs from distinct /24 subnets are necessary for successfully compromising \texttt{ReadRandomNodes}.

\subsection{Exploiting the lookup-buffer}
\label{subsec:findnodeProb}

The lookup-buffer is the second source used by DEVp2p to get potential peers to connect to.
It is populated with a Kademlia-like iterative lookup of a random target ID.
To this end, the local node sends FindNode-Requests with a random target to those nodes from the discovery table that are closest to that target (cf. \Cref{subsec:findnode}).
From the received node set, the 16 closest nodes are used to populate the lookup-buffer, sorted by their distance to the target.
DEVp2p then partially fills the open outbound connections slots by going through the lookup-buffer from the top (\ie minimum distance).

To fill the lookup-buffer with adversarial nodes two steps are necessary:
First, an adversarial node must be queried during the lookup-process.
Second, the node IDs returned by the adversary must be smaller than all other node IDs returned during the lookup.
The first step is \emph{always} given when there is an adversarial node in each bucket: the xor-distance is mainly influenced by the length of the common prefix and each bucket stores node IDs with a specific common prefix length.
Therefore, an adversarial node in each bucket ensures that the attacker is always queried during a lookup (cf. \Cref{subsec:findnode_probability} for a detailed analysis).

The second step can easily be solved by generating sufficiently many node IDs.
Since node IDs are hashed ECDSA keys, they are uniformly distributed over the ID-space; hence, the more node IDs we generate, the higher the chances to be smaller than the rest of the returned IDs.
In the end, by choosing the number of pre-computed keys high enough, it is very likely that all of our $16$ closest IDs are closer to the target than any ID naturally occurring in the Ethereum network.

\section{Analysis of the False Friends Attack}
\label{sec:analysis}

In the following we analyze the mechanics of our false friends attack.
We compute the expected number of necessary key pair generations and for entering every bucket and to exploit the lookup-buffer.
Furthermore, we study FindNode-Requests in detail.

\subsection{Entering a Bucket}

Our false friend attack requires one adversarial node in each bucket.
The question arises how many key pairs we have to generate and how much time it takes to do so.

Since SHA256 is a cryptographic hash function, we assume node IDs to be uniformly distributed~\cite{DBLP:books/daglib/0024293}, \ie, each bit has a probability of $\frac{1}{2}$ of being $0$ or $1$.
Therefore, each ECDSA key pair generation with subsequent hashing corresponds to fair coin tosses repeated independently of each other.
Let $\bar{N}$ be the hashed node ID of the victim node, \ie, $\bar{N}$ is fixed.
Then, for some generated hashed node ID, say $\bar{h}$, the probability that the first bit of $\bar{h}$ is equal to the first bit of $\bar{N}$ is $\frac{1}{2}$.
Recall that the log-distance metric measures the length of the common prefix between $\bar{N}$ and $\bar{h}$ (cf. \Cref{subsec:tableBucketsDist}).
Hence, with probability $\frac{1}{2}$, the two hashes differ at the first bit, which corresponds to a log-distance of 255.
For subsequent buckets the concept is similar: the probability to have a log-distance of 254 is $\frac{1}{4}$ since we have to be equal in the first \emph{and} second bit, \ie, $\frac{1}{2} \cdot \frac{1}{2} = \frac{1}{4}$.
In summary, the probability to have a specific log-distance to a given target hash is 
\begin{align}
p := \Pb[\text{log-distance}(h_1, h_2) = i] = 2^{i-256}.
\label{eq:colProb}
\end{align}

Changing perspective, we can now calculate the expected number of necessary key pair generations to fall into a specific bucket.
Finding a key pair for a desired bucket, or equivalently, a desired log-distance, can be modeled as a series of independent Bernoulli trials until the first success.
Each key pair generation is a Bernoulli trial with success probability $p$ (from \Cref{eq:colProb}).
Repeatedly performing Bernoulli trials and stopping at the first success yields a geometric distribution.
Therefore, generating key pairs for a specific bucket can be modeled as a geometric distribution, with expectation
\begin{align}
\Ex[\text{\# key pair generations for log-distance }i] = \frac{1}{p} = 2^{256-i}.
  \label{eq:expectedNumberGenerations}
\end{align}
For example, to generate an ID with the (in Ethereum) lowest possible log-distance of 239 one would, on average, need $2^{17} = 131072$ key pair generations and hash operations.
Generating a node ID for every bucket requires an average number of operations of
\begin{align}
  \sum_{i=239}^{255} 2^{256-i} = \sum_{i=1}^{17} = 2^i = 262142.
\end{align}
Note that these node ID generations need to be performed only once per victim node.

\subsection{Computing the Probability to Receive a FindNode-Request}
\label{subsec:findnode_probability}

In the following we analyze how probable it is to be asked during a FindNode-Request round.
We distinguish two cases; first, the current implementation in Geth and second, when the buckets would hold more than $k = 16$ nodes.

\subsubsection{Implementation in Geth}

We can assume that node IDs are uniformly distributed in $\{0, 1, \ldots, 2^{256}-1\}$ since they are hashed public keys with 256 bit length.
As in Kademlia, each ID can be viewed as the leaf of a binary tree, thus, each bucket stores node IDs from a specific sub-tree.
For the lookup-process, the victim generates a random target ID, say $D$, and computes its 16 closest neighbors to $D$.
Note, that the number of closest neighbors that Geth computes coincides with maximum possible number of nodes in a bucket.

``Close'' is defined in terms of the simple xor-metric; for two IDs $a, b$ the distance is defined as $d(a, b) := a \oplus b$, taken as integer.
Under the xor-metric, node IDs that have a longer common prefix $D$ are thus considered closer than ones with a shorter common prefix.
Each bucket partitions the binary tree of node IDs into branches by their common prefix.
Therefore, the closest neighbors to $D$ are the ones in $D$'s bucket.

Recall from the previous section that we insert an adversarial node ID into each bucket for our false friends attack.
Hence, searching for the 16 closest neighbors will \emph{always} return at least one attacker-controlled node ID.

\subsubsection{Situation with Larger Buckets}

A simple countermeasure to the current situation in Geth is to simply increase the size of the buckets.
In the following we analyze the probability for an attacker to receive a FindNode-Request in different scenarios.

Without loss of generality assume node IDs to be within $[0, 1]$, as a simplification assume them to be continuously distributed on said interval.
Let $D \in [0, 1]$ be the uniformly random target ID chosen by the lookup-process and $Y_1, \ldots, Y_N$ be the IDs of honest nodes stored in the bucket of $D$, say $b$.
We consider the case $N > 16$.

Assume for the moment that the adversary has exactly \emph{one} node ID in $b$, with ID $Z$.
Although the $Y_1, \ldots, Y_N$ share a common prefix, their suffix is distributed uniformly at random because node IDs are hashes.
For an attacker to receive a FindNode-Request it suffices to be smaller (w.r.t. $D$) than the $17$-th closest ID, \ie, the $17$-th \emph{order statistic}.
We denote the ordering of IDs $Y_i$ with respect to $D$ as $Y_{(1)} <_D Y_{(2)} <_D \ldots <_D Y_{(N)}$, where $Y_{(1)}$ is the node with minimum distance.
It now remains to compute the following probability:
\begin{align}
  \Pb[Z < Y_{(17)}]\label{eq:getAsked}.
\end{align}

The density of $Z$ is $f_Z(z) = 1$, due to its uniformity.
Similarly, let $f_Y(y)$ denote the density of some $Y$.
By definition of probability and the expectation we have for any independent $Y, Z$ with $Z \sim U[0, 1]$:
\begin{align}
  \Pb[Z < Y] &= \int_{y=0}^{1} \underbrace{\int_{z=0}^{y} f_Z(z) \mathrm{d}z}_{=y} f_Y(y) \mathrm{d}y\\
             &= \int_{y=0}^{1} y \cdot f_Y(y) \mathrm{d}y\\
             &= \Ex[Y]. \label{eq:expected}
\end{align}

It is well-known that the order statistics of uniform variables are Beta-distributed~\cite{gentle2009computational}, \ie, $Y_{(l)} \sim Beta(l, N + 1 - l)$.
Inserting that into Equation~\ref{eq:expected} we get
\begin{align}
  \Pb[Z < Y_{(l)}] \overset{(\ref{eq:expected})}{=} \Ex[Y_{(l)}] \overset{\text{Beta distr.}}{=} \frac{l}{N + 1}.
  \label{eq:beta}
\end{align}

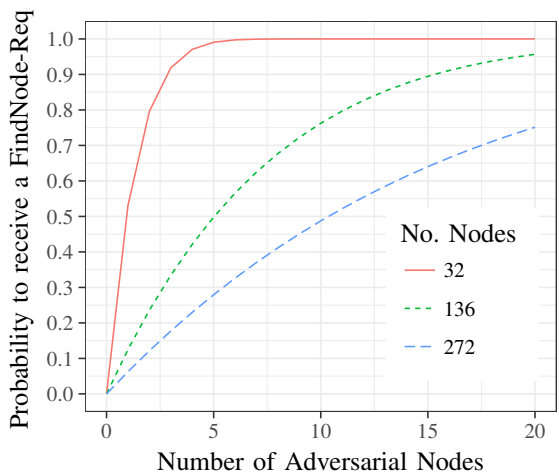
\begin{figure}
  \centering
\begin{tikzpicture}[x=1pt,y=1pt]
\definecolor{fillColor}{RGB}{255,255,255}
\path[use as bounding box,fill=fillColor,fill opacity=0.00] (0,0) rectangle (216.81,180.67);
\begin{scope}
\path[clip] (  0.00,  0.00) rectangle (216.81,180.67);
\definecolor{drawColor}{RGB}{255,255,255}
\definecolor{fillColor}{RGB}{255,255,255}

\path[draw=drawColor,line width= 0.6pt,line join=round,line cap=round,fill=fillColor] ( -0.00,  0.00) rectangle (216.81,180.68);
\end{scope}
\begin{scope}
\path[clip] ( 33.55, 27.87) rectangle (211.81,175.67);
\definecolor{fillColor}{RGB}{255,255,255}

\path[fill=fillColor] ( 33.55, 27.87) rectangle (211.81,175.68);
\definecolor{drawColor}{gray}{0.92}

\path[draw=drawColor,line width= 0.3pt,line join=round] ( 33.55, 27.87) --
  (211.81, 27.87);

\path[draw=drawColor,line width= 0.3pt,line join=round] ( 33.55, 41.31) --
  (211.81, 41.31);

\path[draw=drawColor,line width= 0.3pt,line join=round] ( 33.55, 54.74) --
  (211.81, 54.74);

\path[draw=drawColor,line width= 0.3pt,line join=round] ( 33.55, 68.18) --
  (211.81, 68.18);

\path[draw=drawColor,line width= 0.3pt,line join=round] ( 33.55, 81.62) --
  (211.81, 81.62);

\path[draw=drawColor,line width= 0.3pt,line join=round] ( 33.55, 95.05) --
  (211.81, 95.05);

\path[draw=drawColor,line width= 0.3pt,line join=round] ( 33.55,108.49) --
  (211.81,108.49);

\path[draw=drawColor,line width= 0.3pt,line join=round] ( 33.55,121.93) --
  (211.81,121.93);

\path[draw=drawColor,line width= 0.3pt,line join=round] ( 33.55,135.36) --
  (211.81,135.36);

\path[draw=drawColor,line width= 0.3pt,line join=round] ( 33.55,148.80) --
  (211.81,148.80);

\path[draw=drawColor,line width= 0.3pt,line join=round] ( 33.55,162.24) --
  (211.81,162.24);

\path[draw=drawColor,line width= 0.3pt,line join=round] ( 33.55,175.67) --
  (211.81,175.67);

\path[draw=drawColor,line width= 0.3pt,line join=round] ( 61.91, 27.87) --
  ( 61.91,175.67);

\path[draw=drawColor,line width= 0.3pt,line join=round] (102.42, 27.87) --
  (102.42,175.67);

\path[draw=drawColor,line width= 0.3pt,line join=round] (142.94, 27.87) --
  (142.94,175.67);

\path[draw=drawColor,line width= 0.3pt,line join=round] (183.45, 27.87) --
  (183.45,175.67);

\path[draw=drawColor,line width= 0.6pt,line join=round] ( 33.55, 34.59) --
  (211.81, 34.59);

\path[draw=drawColor,line width= 0.6pt,line join=round] ( 33.55, 48.02) --
  (211.81, 48.02);

\path[draw=drawColor,line width= 0.6pt,line join=round] ( 33.55, 61.46) --
  (211.81, 61.46);

\path[draw=drawColor,line width= 0.6pt,line join=round] ( 33.55, 74.90) --
  (211.81, 74.90);

\path[draw=drawColor,line width= 0.6pt,line join=round] ( 33.55, 88.34) --
  (211.81, 88.34);

\path[draw=drawColor,line width= 0.6pt,line join=round] ( 33.55,101.77) --
  (211.81,101.77);

\path[draw=drawColor,line width= 0.6pt,line join=round] ( 33.55,115.21) --
  (211.81,115.21);

\path[draw=drawColor,line width= 0.6pt,line join=round] ( 33.55,128.65) --
  (211.81,128.65);

\path[draw=drawColor,line width= 0.6pt,line join=round] ( 33.55,142.08) --
  (211.81,142.08);

\path[draw=drawColor,line width= 0.6pt,line join=round] ( 33.55,155.52) --
  (211.81,155.52);

\path[draw=drawColor,line width= 0.6pt,line join=round] ( 33.55,168.96) --
  (211.81,168.96);

\path[draw=drawColor,line width= 0.6pt,line join=round] ( 41.65, 27.87) --
  ( 41.65,175.67);

\path[draw=drawColor,line width= 0.6pt,line join=round] ( 82.17, 27.87) --
  ( 82.17,175.67);

\path[draw=drawColor,line width= 0.6pt,line join=round] (122.68, 27.87) --
  (122.68,175.67);

\path[draw=drawColor,line width= 0.6pt,line join=round] (163.19, 27.87) --
  (163.19,175.67);

\path[draw=drawColor,line width= 0.6pt,line join=round] (203.71, 27.87) --
  (203.71,175.67);
\definecolor{drawColor}{RGB}{248,118,109}

\path[draw=drawColor,line width= 0.6pt,line join=round] ( 41.65, 34.59) --
  ( 49.76,105.97) --
  ( 57.86,141.55) --
  ( 65.96,158.02) --
  ( 74.06,165.02) --
  ( 82.17,167.70) --
  ( 90.27,168.61) --
  ( 98.37,168.88) --
  (106.48,168.94) --
  (114.58,168.95) --
  (122.68,168.96) --
  (130.78,168.96) --
  (138.89,168.96) --
  (146.99,168.96) --
  (155.09,168.96) --
  (163.19,168.96) --
  (171.30,168.96) --
  (179.40,168.96) --
  (187.50,168.96) --
  (195.60,168.96) --
  (203.71,168.96);
\definecolor{drawColor}{RGB}{0,186,56}

\path[draw=drawColor,line width= 0.6pt,dash pattern=on 2pt off 2pt ,line join=round] ( 41.65, 34.59) --
  ( 49.76, 51.38) --
  ( 57.86, 66.30) --
  ( 65.96, 79.51) --
  ( 74.06, 91.20) --
  ( 82.17,101.52) --
  ( 90.27,110.60) --
  ( 98.37,118.58) --
  (106.48,125.57) --
  (114.58,131.69) --
  (122.68,137.03) --
  (130.78,141.67) --
  (138.89,145.70) --
  (146.99,149.20) --
  (155.09,152.21) --
  (163.19,154.81) --
  (171.30,157.04) --
  (179.40,158.95) --
  (187.50,160.58) --
  (195.60,161.96) --
  (203.71,163.14);
\definecolor{drawColor}{RGB}{97,156,255}

\path[draw=drawColor,line width= 0.6pt,dash pattern=on 4pt off 2pt ,line join=round] ( 41.65, 34.59) --
  ( 49.76, 42.99) --
  ( 57.86, 50.92) --
  ( 65.96, 58.40) --
  ( 74.06, 65.47) --
  ( 82.17, 72.13) --
  ( 90.27, 78.41) --
  ( 98.37, 84.33) --
  (106.48, 89.90) --
  (114.58, 95.14) --
  (122.68,100.08) --
  (130.78,104.72) --
  (138.89,109.08) --
  (146.99,113.17) --
  (155.09,117.01) --
  (163.19,120.62) --
  (171.30,124.00) --
  (179.40,127.18) --
  (187.50,130.14) --
  (195.60,132.93) --
  (203.71,135.53);
\definecolor{drawColor}{gray}{0.20}

\path[draw=drawColor,line width= 0.6pt,line join=round,line cap=round] ( 33.55, 27.87) rectangle (211.81,175.68);
\end{scope}
\begin{scope}
\path[clip] (  0.00,  0.00) rectangle (216.81,180.67);
\definecolor{drawColor}{gray}{0.30}

\node[text=drawColor,anchor=base east,inner sep=0pt, outer sep=0pt, scale=  0.80] at ( 29.05, 31.83) {0.0};

\node[text=drawColor,anchor=base east,inner sep=0pt, outer sep=0pt, scale=  0.80] at ( 29.05, 45.27) {0.1};

\node[text=drawColor,anchor=base east,inner sep=0pt, outer sep=0pt, scale=  0.80] at ( 29.05, 58.71) {0.2};

\node[text=drawColor,anchor=base east,inner sep=0pt, outer sep=0pt, scale=  0.80] at ( 29.05, 72.14) {0.3};

\node[text=drawColor,anchor=base east,inner sep=0pt, outer sep=0pt, scale=  0.80] at ( 29.05, 85.58) {0.4};

\node[text=drawColor,anchor=base east,inner sep=0pt, outer sep=0pt, scale=  0.80] at ( 29.05, 99.02) {0.5};

\node[text=drawColor,anchor=base east,inner sep=0pt, outer sep=0pt, scale=  0.80] at ( 29.05,112.45) {0.6};

\node[text=drawColor,anchor=base east,inner sep=0pt, outer sep=0pt, scale=  0.80] at ( 29.05,125.89) {0.7};

\node[text=drawColor,anchor=base east,inner sep=0pt, outer sep=0pt, scale=  0.80] at ( 29.05,139.33) {0.8};

\node[text=drawColor,anchor=base east,inner sep=0pt, outer sep=0pt, scale=  0.80] at ( 29.05,152.76) {0.9};

\node[text=drawColor,anchor=base east,inner sep=0pt, outer sep=0pt, scale=  0.80] at ( 29.05,166.20) {1.0};
\end{scope}
\begin{scope}
\path[clip] (  0.00,  0.00) rectangle (216.81,180.67);
\definecolor{drawColor}{gray}{0.20}

\path[draw=drawColor,line width= 0.6pt,line join=round] ( 31.05, 34.59) --
  ( 33.55, 34.59);

\path[draw=drawColor,line width= 0.6pt,line join=round] ( 31.05, 48.02) --
  ( 33.55, 48.02);

\path[draw=drawColor,line width= 0.6pt,line join=round] ( 31.05, 61.46) --
  ( 33.55, 61.46);

\path[draw=drawColor,line width= 0.6pt,line join=round] ( 31.05, 74.90) --
  ( 33.55, 74.90);

\path[draw=drawColor,line width= 0.6pt,line join=round] ( 31.05, 88.34) --
  ( 33.55, 88.34);

\path[draw=drawColor,line width= 0.6pt,line join=round] ( 31.05,101.77) --
  ( 33.55,101.77);

\path[draw=drawColor,line width= 0.6pt,line join=round] ( 31.05,115.21) --
  ( 33.55,115.21);

\path[draw=drawColor,line width= 0.6pt,line join=round] ( 31.05,128.65) --
  ( 33.55,128.65);

\path[draw=drawColor,line width= 0.6pt,line join=round] ( 31.05,142.08) --
  ( 33.55,142.08);

\path[draw=drawColor,line width= 0.6pt,line join=round] ( 31.05,155.52) --
  ( 33.55,155.52);

\path[draw=drawColor,line width= 0.6pt,line join=round] ( 31.05,168.96) --
  ( 33.55,168.96);
\end{scope}
\begin{scope}
\path[clip] (  0.00,  0.00) rectangle (216.81,180.67);
\definecolor{drawColor}{gray}{0.20}

\path[draw=drawColor,line width= 0.6pt,line join=round] ( 41.65, 25.37) --
  ( 41.65, 27.87);

\path[draw=drawColor,line width= 0.6pt,line join=round] ( 82.17, 25.37) --
  ( 82.17, 27.87);

\path[draw=drawColor,line width= 0.6pt,line join=round] (122.68, 25.37) --
  (122.68, 27.87);

\path[draw=drawColor,line width= 0.6pt,line join=round] (163.19, 25.37) --
  (163.19, 27.87);

\path[draw=drawColor,line width= 0.6pt,line join=round] (203.71, 25.37) --
  (203.71, 27.87);
\end{scope}
\begin{scope}
\path[clip] (  0.00,  0.00) rectangle (216.81,180.67);
\definecolor{drawColor}{gray}{0.30}

\node[text=drawColor,anchor=base,inner sep=0pt, outer sep=0pt, scale=  0.80] at ( 41.65, 17.86) {0};

\node[text=drawColor,anchor=base,inner sep=0pt, outer sep=0pt, scale=  0.80] at ( 82.17, 17.86) {5};

\node[text=drawColor,anchor=base,inner sep=0pt, outer sep=0pt, scale=  0.80] at (122.68, 17.86) {10};

\node[text=drawColor,anchor=base,inner sep=0pt, outer sep=0pt, scale=  0.80] at (163.19, 17.86) {15};

\node[text=drawColor,anchor=base,inner sep=0pt, outer sep=0pt, scale=  0.80] at (203.71, 17.86) {20};
\end{scope}
\begin{scope}
\path[clip] (  0.00,  0.00) rectangle (216.81,180.67);
\definecolor{drawColor}{RGB}{0,0,0}

\node[text=drawColor,anchor=base,inner sep=0pt, outer sep=0pt, scale=  1.00] at (122.68,  5.97) {Number of Adversarial Nodes};
\end{scope}
\begin{scope}
\path[clip] (  0.00,  0.00) rectangle (216.81,180.67);
\definecolor{drawColor}{RGB}{0,0,0}

\node[text=drawColor,rotate= 90.00,anchor=base,inner sep=0pt, outer sep=0pt, scale=  1.00] at ( 11.89,101.77) {Probability to receive a FindNode-Req.};
\end{scope}
\begin{scope}
\path[clip] (  0.00,  0.00) rectangle (216.81,180.67);
\definecolor{fillColor}{RGB}{255,255,255}

\path[fill=fillColor] (147.25, 39.59) rectangle (205.07,104.83);
\end{scope}
\begin{scope}
\path[clip] (  0.00,  0.00) rectangle (216.81,180.67);
\definecolor{drawColor}{RGB}{0,0,0}

\node[text=drawColor,anchor=base west,inner sep=0pt, outer sep=0pt, scale=  1.00] at (152.94, 92.25) {No. Nodes};
\end{scope}
\begin{scope}
\path[clip] (  0.00,  0.00) rectangle (216.81,180.67);
\definecolor{fillColor}{RGB}{255,255,255}

\path[fill=fillColor] (152.94, 74.19) rectangle (167.40, 88.64);
\end{scope}
\begin{scope}
\path[clip] (  0.00,  0.00) rectangle (216.81,180.67);
\definecolor{drawColor}{RGB}{248,118,109}

\path[draw=drawColor,line width= 0.6pt,line join=round] (154.39, 81.41) -- (165.95, 81.41);
\end{scope}
\begin{scope}
\path[clip] (  0.00,  0.00) rectangle (216.81,180.67);
\definecolor{fillColor}{RGB}{255,255,255}

\path[fill=fillColor] (152.94, 59.73) rectangle (167.40, 74.19);
\end{scope}
\begin{scope}
\path[clip] (  0.00,  0.00) rectangle (216.81,180.67);
\definecolor{drawColor}{RGB}{0,186,56}

\path[draw=drawColor,line width= 0.6pt,dash pattern=on 2pt off 2pt ,line join=round] (154.39, 66.96) -- (165.95, 66.96);
\end{scope}
\begin{scope}
\path[clip] (  0.00,  0.00) rectangle (216.81,180.67);
\definecolor{fillColor}{RGB}{255,255,255}

\path[fill=fillColor] (152.94, 45.28) rectangle (167.40, 59.73);
\end{scope}
\begin{scope}
\path[clip] (  0.00,  0.00) rectangle (216.81,180.67);
\definecolor{drawColor}{RGB}{97,156,255}

\path[draw=drawColor,line width= 0.6pt,dash pattern=on 4pt off 2pt ,line join=round] (154.39, 52.51) -- (165.95, 52.51);
\end{scope}
\begin{scope}
\path[clip] (  0.00,  0.00) rectangle (216.81,180.67);
\definecolor{drawColor}{RGB}{0,0,0}

\node[text=drawColor,anchor=base west,inner sep=0pt, outer sep=0pt, scale=  0.80] at (169.20, 78.66) {32};
\end{scope}
\begin{scope}
\path[clip] (  0.00,  0.00) rectangle (216.81,180.67);
\definecolor{drawColor}{RGB}{0,0,0}

\node[text=drawColor,anchor=base west,inner sep=0pt, outer sep=0pt, scale=  0.80] at (169.20, 64.21) {136};
\end{scope}
\begin{scope}
\path[clip] (  0.00,  0.00) rectangle (216.81,180.67);
\definecolor{drawColor}{RGB}{0,0,0}

\node[text=drawColor,anchor=base west,inner sep=0pt, outer sep=0pt, scale=  0.80] at (169.20, 49.75) {272};
\end{scope}
\end{tikzpicture}

  \caption{Probability that the adversary gets queried with a FindNode-Request for a given number of adversarial nodes in the victims discovery table.}
  \label{fig:findnode}
\end{figure}

In general, the attacker can have multiple, say $a \in \N$ nodes in the bucket $b$.
To get queried, \emph{at least one attacker ID} has to be within the closest nodes.
Denote the adversarial IDs by $Z_1, \ldots, Z_a \sim U[0, 1]$.
Then we obtain:
\begin{align}
  &\Pb[\text{At least one attacker ID within } k \text{ closest}]\\
  &= 1 - \Pb[Z_1 > Y_{(l)} \land  Z_2 > Y_{(l)} \land \ldots Z_a > Y_{(l)}]\\
  &\overset{i.i.d.}{=} 1 - \Pb[Z_1 > Y_{(l)}] \cdot \ldots \cdot \Pb[Z_a > Y_{(l)}]\\
  &\overset{(\ref{eq:beta})}{=} 1 - \left[ 1 - (\frac{l}{N + 1})\right]^a. \label{eq:finalFindnode}
\end{align}
Intuitively, the more adversarial nodes there are in bucket $b$, the more unlikely it becomes \emph{not} to get queried during the lookup-process.
\Cref{fig:findnode} shows the result of \Cref{eq:finalFindnode}.
The probability for the adversary to receive a FindNode-Request is plotted over the number of adversarial nodes in bucket $b$.
For the number of nodes per bucket we consider three cases:
\begin{enumerate}
  \item $N = 32$, double the size of current buckets.
  \item $N = 136$, \ie, the size of a bucket corresponds to half of the current size of the complete discovery table.
  \item $N = 272$ which corresponds to the maximum size of the current discovery table (17 buckets à 16 nodes each).
\end{enumerate}

\subsection{Filling the Lookup-buffer with Pre-Computed Node IDs}

Recall, that in order to take over the lookup-buffer, we identified two necessary steps:
First, an adversarial node must be queried during the lookup-process.
Although this is purely a matter of chance, we saw that the probability to get queried is relatively high even with a small number of nodes.
In a second step, the node IDs returned by the adversary must be smaller (with respect to the random target) than all other node IDs returned during the lookup.
Intuitively, this can be ensured by pre-computing a large number of node IDs, since each ID generation corresponds to a draw from the uniform distribution.
Every draw has the same probability to be smaller than any other node ID on the Ethereum network.
Therefore, the more node IDs we generate, the more likely this event becomes.
The question that remains is the following: how many ECDSA key pairs should an adversary generate in advance in order to almost always return the smallest node ID?

To model this scenario, assume all other nodes already replied to the FindNode-Request with target $D$, which yields node IDs (sorted w.l.o.g) $0 \leq x_1, < x_2 \ldots < x_m \leq 1$.
\footnote{Technically, the sorting is w.r.t. to $<_D$. The xor-operation only induces a permutation of the $x_i$, preserving the uniform distribution. We therefore purposely omit the xor-operation for the ease of understanding.}
Each node ID induces two intervals around itself on the space of possible node IDs:
\begin{align}
  [0, x_1), (x_1, x_2), (x_2, x_3), \ldots, (x_m, 1].
\end{align}
For a new node ID to be the minimum, it has to fall into the first interval and only the first interval.
Each interval is equally likely to occur, because the likelihood of falling into an interval only depends on its size when sampling from a uniform distribution.
Since the existing node IDs $x_1, \ldots, x_m$ are uniformly distributed themselves, the intervals have, on average, the same size.

Let $Y \sim U[0,1]$ represent a node ID generation.
There are $m + 1$ intervals, the chances of falling into the first interval (\ie, having the minimum node ID) by uniform sampling is
\begin{align}
  \Pb[Y <_D \min\{X_1, \ldots, X_m\}] = \frac{1}{m+1}.
\end{align}

In other words, every generated node ID has a chance of $p := \frac{1}{m+1}$ of being the minimum node ID.
Therefore, repeating the process of generating node IDs again yields a Bernoulli trial with success probability $p$.

\begin{figure}
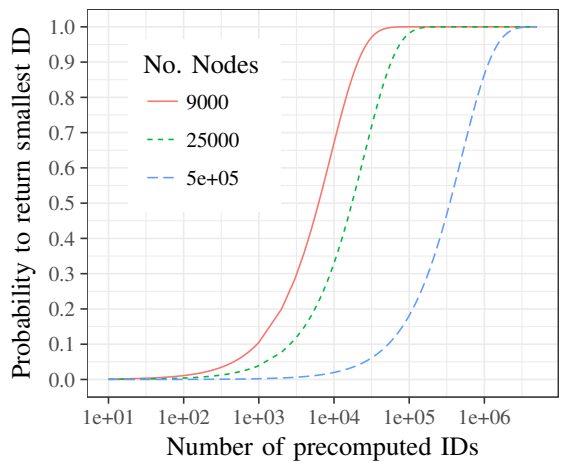

  \centering


  \caption{Probability that the lowest node ID is returned, depending on the number of honest nodes in the network.}
  \label{fig:smallestIDProb}
\end{figure}

In reality, we do not know the other node IDs before we start generating our own, meaning that whether a draw was successful cannot be determined.
Instead, an adversary pre-computes a large number of IDs and simply returns the smallest ones with respect to the random target, if she receives a FindNode-Request.
Still, we can bound the probability that \emph{at least one} of our draws is smaller than the minimum returned by the honest nodes.
Let there be $m$ node IDs in the Ethereum network and $n$ pre-computed node IDs by the adversary.
For convenience, we define $Y_{\min} := \min\{Y_1, \ldots, Y_n\}$ and $X_{\min} := \min\{X_1, \ldots, X_m\}$.

\begin{align}
  \Pb[Y_{\min} <_D X_{\min}] = 1 - (1 - \frac{1}{m + 1})^n
\end{align}
Naturally, this is a lower bound on the probability, since a lookup normally does not yield the minimum node ID of the complete Ethereum network.
The resulting probability is depicted in \Cref{fig:smallestIDProb} for different choices of $n$, the number of node IDs in the network.
It can be seen that the more honest nodes there are, the smaller the probability for an adversary to have the minimal node ID becomes.

Note that we have to distinguish between node IDs in the discovery table and actual nodes on the main Ethereum network.
The Ethereum protocol is running concurrently with other protocols on the same communication channels and packet structures; therefore the number of node IDs is ten times larger than the number of Ethereum nodes at roughly $3 \cdot 10^6$ node IDs~\cite{DBLP:conf/imc/KimMMMMB18}.
In case of returning the smallest ID for a FindNode-Request, this behavior slightly raises the bar for an attacker.

The red and green lines depict a situation where every node ID would correspond to an actual node on the Ethereum network.
The red line shows the probability for $n = 9000$ nodes, as reported by Ethernode\footnote{\url{https://www.ethernodes.org/network/1}}.
The green line corresponds to $n = 25000$ nodes in the network, the sum of all approaches discussed in~\cite{DBLP:conf/imc/KimMMMMB18}.
The blue line corresponds to an upper bound on the number of node IDs of $n = 5\cdot10^5$ nodes.
In all cases, however, $5\cdot10^6$ pre-computed ECDSA key-pairs are enough to return the minimal node ID almost certainly.
\section{Evaluation}
\label{sec:eval}
We evaluated the previously described concepts using a victim node deployed specifically for this task.
The victim had the latest Geth version from github (v1.8.20) and was connected to the Ethereum main network.
\subsection{Pre-Computing Node IDs}
\begin{figure}
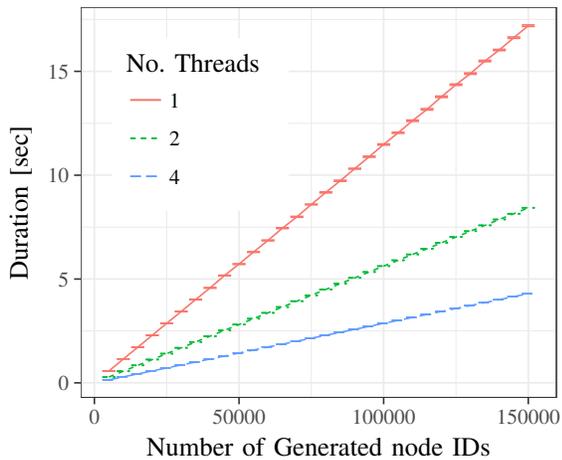

\centering


  \caption{Mean duration of node ID generation with different numbers of parallel threads.
  The (very small) error bars show the 95\% confidence interval for 100 runs.}
\label{fig:nodeidBench}
\end{figure}

The generation of node IDs is essential for placing a node in each bucket and to ensure that the lowest ID is returned during the lookup-process.
Therefore, we measured the calculation time for new ECDSA key pairs and the corresponding hashes.
Our measurements were conducted on a system with an Intel\textsuperscript{\textregistered} Core\texttrademark{} i5-6600K processor with four logical cores.
The results depicted in \Cref{fig:nodeidBench} show that, on average, 35000 ECDSA keys and corresponding hashes can be generated per second when using four parallel threads.
Generating a node ID to enter bucket number 239 (the smallest bucket) would therefore only take \SI{7.5}{\second} on average.
For the real-world implementation of the attack, we pre-computed $5\cdot10^6$ node IDs to hijack the lookup-buffer.
Computing this many IDs takes roughly \SI{2.5}{\minute} using four parallel threads and \SI{11}{\minute} with a single thread.
Note that even when attacking different victim nodes this computation has to be performed only once, since the target of a lookup is random and does not depend on any victim-specific information.
\subsection{Attack Implementation}%

First, to compare the performance of the attack to~\cite{DBLP:journals/iacr/MarcusHG18}, we repeatedly attacked a recently restarted victim.
Recall that we do not necessarily require that the victim is restarted, since our attack relies on high peer churn.
The reliance on churn implies a delay, as previously established connections must be terminated before their slots can be occupied by an adversary.
However, the waiting time is small for recently restarted nodes.

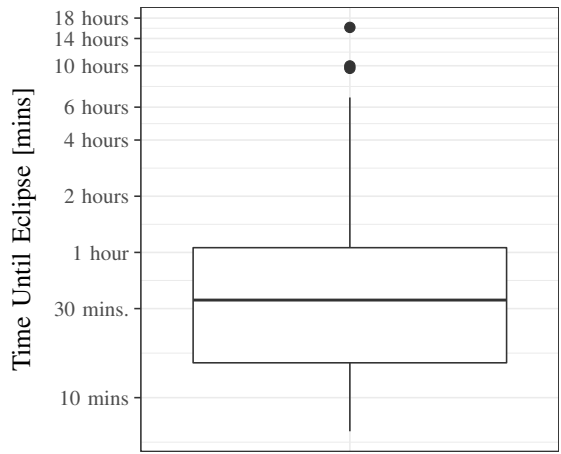
\begin{figure}
\begin{tikzpicture}[x=1pt,y=1pt]
\definecolor{fillColor}{RGB}{255,255,255}
\path[use as bounding box,fill=fillColor,fill opacity=0.00] (0,0) rectangle (216.81,180.67);
\begin{scope}
\path[clip] (  0.00,  0.00) rectangle (216.81,180.67);
\definecolor{drawColor}{RGB}{255,255,255}
\definecolor{fillColor}{RGB}{255,255,255}

\path[draw=drawColor,line width= 0.6pt,line join=round,line cap=round,fill=fillColor] (  0.00,  0.00) rectangle (216.81,180.68);
\end{scope}
\begin{scope}
\path[clip] ( 53.72,  7.50) rectangle (211.81,175.67);
\definecolor{fillColor}{RGB}{255,255,255}

\path[fill=fillColor] ( 53.72,  7.50) rectangle (211.81,175.68);
\definecolor{drawColor}{gray}{0.92}

\path[draw=drawColor,line width= 0.3pt,line join=round] ( 53.72, 11.02) --
  (211.81, 11.02);

\path[draw=drawColor,line width= 0.3pt,line join=round] ( 53.72, 44.73) --
  (211.81, 44.73);

\path[draw=drawColor,line width= 0.3pt,line join=round] ( 53.72, 72.22) --
  (211.81, 72.22);

\path[draw=drawColor,line width= 0.3pt,line join=round] ( 53.72, 93.49) --
  (211.81, 93.49);

\path[draw=drawColor,line width= 0.3pt,line join=round] ( 53.72,114.76) --
  (211.81,114.76);

\path[draw=drawColor,line width= 0.3pt,line join=round] ( 53.72,131.61) --
  (211.81,131.61);

\path[draw=drawColor,line width= 0.3pt,line join=round] ( 53.72,145.67) --
  (211.81,145.67);

\path[draw=drawColor,line width= 0.3pt,line join=round] ( 53.72,158.67) --
  (211.81,158.67);

\path[draw=drawColor,line width= 0.3pt,line join=round] ( 53.72,167.69) --
  (211.81,167.69);

\path[draw=drawColor,line width= 0.6pt,line join=round] ( 53.72, 27.87) --
  (211.81, 27.87);

\path[draw=drawColor,line width= 0.6pt,line join=round] ( 53.72, 61.58) --
  (211.81, 61.58);

\path[draw=drawColor,line width= 0.6pt,line join=round] ( 53.72, 82.85) --
  (211.81, 82.85);

\path[draw=drawColor,line width= 0.6pt,line join=round] ( 53.72,104.12) --
  (211.81,104.12);

\path[draw=drawColor,line width= 0.6pt,line join=round] ( 53.72,125.39) --
  (211.81,125.39);

\path[draw=drawColor,line width= 0.6pt,line join=round] ( 53.72,137.84) --
  (211.81,137.84);

\path[draw=drawColor,line width= 0.6pt,line join=round] ( 53.72,153.51) --
  (211.81,153.51);

\path[draw=drawColor,line width= 0.6pt,line join=round] ( 53.72,163.83) --
  (211.81,163.83);

\path[draw=drawColor,line width= 0.6pt,line join=round] ( 53.72,171.55) --
  (211.81,171.55);

\path[draw=drawColor,line width= 0.6pt,line join=round] (132.77,  7.50) --
  (132.77,175.67);
\definecolor{drawColor}{gray}{0.20}
\definecolor{fillColor}{gray}{0.20}

\path[draw=drawColor,line width= 0.4pt,line join=round,line cap=round,fill=fillColor] (132.77,152.53) circle (  1.96);

\path[draw=drawColor,line width= 0.4pt,line join=round,line cap=round,fill=fillColor] (132.77,153.44) circle (  1.96);

\path[draw=drawColor,line width= 0.4pt,line join=round,line cap=round,fill=fillColor] (132.77,168.03) circle (  1.96);

\path[draw=drawColor,line width= 0.6pt,line join=round] (132.77, 84.64) -- (132.77,141.50);

\path[draw=drawColor,line width= 0.6pt,line join=round] (132.77, 41.06) -- (132.77, 15.14);
\definecolor{fillColor}{RGB}{255,255,255}

\path[draw=drawColor,line width= 0.6pt,line join=round,line cap=round,fill=fillColor] ( 73.48, 84.64) --
  ( 73.48, 41.06) --
  (192.05, 41.06) --
  (192.05, 84.64) --
  ( 73.48, 84.64) --
  cycle;

\path[draw=drawColor,line width= 1.1pt,line join=round] ( 73.48, 64.81) -- (192.05, 64.81);

\path[draw=drawColor,line width= 0.6pt,line join=round,line cap=round] ( 53.72,  7.50) rectangle (211.81,175.68);
\end{scope}
\begin{scope}
\path[clip] (  0.00,  0.00) rectangle (216.81,180.67);
\definecolor{drawColor}{gray}{0.30}

\node[text=drawColor,anchor=base east,inner sep=0pt, outer sep=0pt, scale=  0.80] at ( 49.22, 25.12) {10 mins};

\node[text=drawColor,anchor=base east,inner sep=0pt, outer sep=0pt, scale=  0.80] at ( 49.22, 58.83) {30 mins.};

\node[text=drawColor,anchor=base east,inner sep=0pt, outer sep=0pt, scale=  0.80] at ( 49.22, 80.10) {1 hour};

\node[text=drawColor,anchor=base east,inner sep=0pt, outer sep=0pt, scale=  0.80] at ( 49.22,101.37) {2 hours};

\node[text=drawColor,anchor=base east,inner sep=0pt, outer sep=0pt, scale=  0.80] at ( 49.22,122.64) {4 hours};

\node[text=drawColor,anchor=base east,inner sep=0pt, outer sep=0pt, scale=  0.80] at ( 49.22,135.08) {6 hours};

\node[text=drawColor,anchor=base east,inner sep=0pt, outer sep=0pt, scale=  0.80] at ( 49.22,150.76) {10 hours};

\node[text=drawColor,anchor=base east,inner sep=0pt, outer sep=0pt, scale=  0.80] at ( 49.22,161.08) {14 hours};

\node[text=drawColor,anchor=base east,inner sep=0pt, outer sep=0pt, scale=  0.80] at ( 49.22,168.79) {18 hours};
\end{scope}
\begin{scope}
\path[clip] (  0.00,  0.00) rectangle (216.81,180.67);
\definecolor{drawColor}{gray}{0.20}

\path[draw=drawColor,line width= 0.6pt,line join=round] ( 51.22, 27.87) --
  ( 53.72, 27.87);

\path[draw=drawColor,line width= 0.6pt,line join=round] ( 51.22, 61.58) --
  ( 53.72, 61.58);

\path[draw=drawColor,line width= 0.6pt,line join=round] ( 51.22, 82.85) --
  ( 53.72, 82.85);

\path[draw=drawColor,line width= 0.6pt,line join=round] ( 51.22,104.12) --
  ( 53.72,104.12);

\path[draw=drawColor,line width= 0.6pt,line join=round] ( 51.22,125.39) --
  ( 53.72,125.39);

\path[draw=drawColor,line width= 0.6pt,line join=round] ( 51.22,137.84) --
  ( 53.72,137.84);

\path[draw=drawColor,line width= 0.6pt,line join=round] ( 51.22,153.51) --
  ( 53.72,153.51);

\path[draw=drawColor,line width= 0.6pt,line join=round] ( 51.22,163.83) --
  ( 53.72,163.83);

\path[draw=drawColor,line width= 0.6pt,line join=round] ( 51.22,171.55) --
  ( 53.72,171.55);
\end{scope}
\begin{scope}
\path[clip] (  0.00,  0.00) rectangle (216.81,180.67);
\definecolor{drawColor}{RGB}{0,0,0}

\node[text=drawColor,rotate= 90.00,anchor=base,inner sep=0pt, outer sep=0pt, scale=  1.00] at ( 11.89, 91.59) {Time Until Eclipse [mins]};
\end{scope}
\end{tikzpicture}

  \caption{Log-scale box plot of attack durations when attacking a recently started victim. The plot depicts the median duration as well as the upper and lower quartile and all outliers outside 1.5 times the interquartile range.}
  \label{fig:attackDuration}
\end{figure}

The attack was started immediately after the victim node became online.
No connection slot was yet occupied, but the neighbor table always contained benign nodes, due to countermeasures introduced in~\cite{DBLP:journals/iacr/MarcusHG18}.
We measured the time until every slot was filled with an attacker-controlled node, with a cutoff timeout of \SI{24}{\hour} after which the experiment was restarted.
One could argue that also in unsuccessful attempts, the victim would eventually have been eclipsed.
The attack was repeated 50 times, out of which 45 times were successful within the cutoff timeout, whereas 5 attempts did not complete in that time.
\Cref{fig:attackDuration} shows the results in a log-scale box plot.
The box depicts the upper and lower quartile of measured durations, the median is indicated as a solid line inside the box.
Measurements outside 1.5 times the interquartile range are considered as outliers, plotted as dots.

It can be seen that out of the 45 successful attacks \SI{75}{\percent} completed in just over \SI{60}{\minute}, indicating that an adversary can eclipse recently restarted nodes within a reasonable time span.
This finding implies that peer churn in the neighbor table and in the peer management is high in recently restarted nodes.

\subsection{Distribution of Connection Durations}

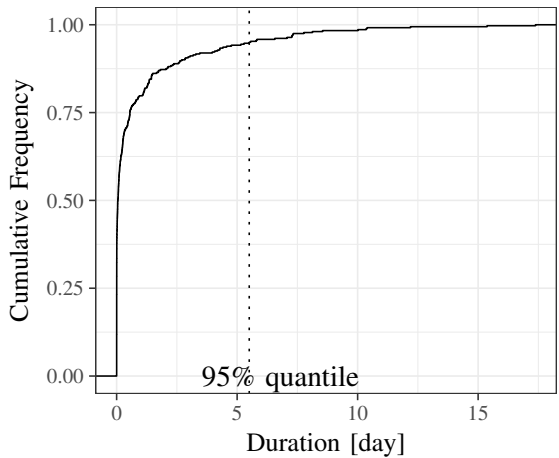
\begin{figure}
  \centering
\begin{tikzpicture}[x=1pt,y=1pt]
\definecolor{fillColor}{RGB}{255,255,255}
\path[use as bounding box,fill=fillColor,fill opacity=0.00] (0,0) rectangle (216.81,180.67);
\begin{scope}
\path[clip] (  0.00,  0.00) rectangle (216.81,180.67);
\definecolor{drawColor}{RGB}{255,255,255}
\definecolor{fillColor}{RGB}{255,255,255}

\path[draw=drawColor,line width= 0.6pt,line join=round,line cap=round,fill=fillColor] ( -0.00,  0.00) rectangle (216.81,180.68);
\end{scope}
\begin{scope}
\path[clip] ( 37.55, 29.40) rectangle (211.81,175.68);
\definecolor{fillColor}{RGB}{255,255,255}

\path[fill=fillColor] ( 37.55, 29.40) rectangle (211.81,175.68);
\definecolor{drawColor}{gray}{0.92}

\path[draw=drawColor,line width= 0.3pt,line join=round] ( 37.55, 52.67) --
  (211.81, 52.67);

\path[draw=drawColor,line width= 0.3pt,line join=round] ( 37.55, 85.91) --
  (211.81, 85.91);

\path[draw=drawColor,line width= 0.3pt,line join=round] ( 37.55,119.16) --
  (211.81,119.16);

\path[draw=drawColor,line width= 0.3pt,line join=round] ( 37.55,152.40) --
  (211.81,152.40);

\path[draw=drawColor,line width= 0.3pt,line join=round] ( 68.25, 29.40) --
  ( 68.25,175.68);

\path[draw=drawColor,line width= 0.3pt,line join=round] (113.83, 29.40) --
  (113.83,175.68);

\path[draw=drawColor,line width= 0.3pt,line join=round] (159.41, 29.40) --
  (159.41,175.68);

\path[draw=drawColor,line width= 0.3pt,line join=round] (204.99, 29.40) --
  (204.99,175.68);

\path[draw=drawColor,line width= 0.6pt,line join=round] ( 37.55, 36.05) --
  (211.81, 36.05);

\path[draw=drawColor,line width= 0.6pt,line join=round] ( 37.55, 69.29) --
  (211.81, 69.29);

\path[draw=drawColor,line width= 0.6pt,line join=round] ( 37.55,102.54) --
  (211.81,102.54);

\path[draw=drawColor,line width= 0.6pt,line join=round] ( 37.55,135.78) --
  (211.81,135.78);

\path[draw=drawColor,line width= 0.6pt,line join=round] ( 37.55,169.03) --
  (211.81,169.03);

\path[draw=drawColor,line width= 0.6pt,line join=round] ( 45.46, 29.40) --
  ( 45.46,175.68);

\path[draw=drawColor,line width= 0.6pt,line join=round] ( 91.04, 29.40) --
  ( 91.04,175.68);

\path[draw=drawColor,line width= 0.6pt,line join=round] (136.62, 29.40) --
  (136.62,175.68);

\path[draw=drawColor,line width= 0.6pt,line join=round] (182.20, 29.40) --
  (182.20,175.68);
\definecolor{drawColor}{RGB}{0,0,0}

\path[draw=drawColor,line width= 0.6pt,line join=round] ( 37.55, 36.05) --
  ( 45.47, 36.05) --
  ( 45.47, 36.41) --
  ( 45.47, 36.41) --
  ( 45.47, 37.52) --
  ( 45.47, 37.52) --
  ( 45.47, 38.26) --
  ( 45.47, 38.26) --
  ( 45.47, 38.99) --
  ( 45.47, 38.99) --
  ( 45.47, 39.36) --
  ( 45.47, 39.36) --
  ( 45.47, 39.73) --
  ( 45.47, 39.73) --
  ( 45.47, 40.10) --
  ( 45.47, 40.10) --
  ( 45.47, 40.83) --
  ( 45.47, 40.83) --
  ( 45.47, 41.94) --
  ( 45.47, 41.94) --
  ( 45.47, 42.31) --
  ( 45.47, 42.31) --
  ( 45.47, 43.04) --
  ( 45.47, 43.04) --
  ( 45.47, 43.41) --
  ( 45.47, 43.41) --
  ( 45.47, 44.52) --
  ( 45.47, 44.52) --
  ( 45.47, 44.89) --
  ( 45.47, 44.89) --
  ( 45.47, 45.62) --
  ( 45.47, 45.62) --
  ( 45.47, 45.99) --
  ( 45.47, 45.99) --
  ( 45.47, 46.36) --
  ( 45.47, 46.36) --
  ( 45.47, 46.73) --
  ( 45.47, 46.73) --
  ( 45.47, 47.46) --
  ( 45.47, 47.46) --
  ( 45.47, 47.83) --
  ( 45.47, 47.83) --
  ( 45.47, 48.20) --
  ( 45.47, 48.20) --
  ( 45.47, 48.57) --
  ( 45.47, 48.57) --
  ( 45.47, 48.94) --
  ( 45.47, 48.94) --
  ( 45.47, 49.31) --
  ( 45.47, 49.31) --
  ( 45.47, 50.04) --
  ( 45.47, 50.04) --
  ( 45.47, 50.78) --
  ( 45.47, 50.78) --
  ( 45.47, 51.89) --
  ( 45.48, 51.89) --
  ( 45.48, 52.25) --
  ( 45.48, 52.25) --
  ( 45.48, 52.62) --
  ( 45.48, 52.62) --
  ( 45.48, 52.99) --
  ( 45.48, 52.99) --
  ( 45.48, 53.36) --
  ( 45.48, 53.36) --
  ( 45.48, 53.73) --
  ( 45.48, 53.73) --
  ( 45.48, 54.46) --
  ( 45.48, 54.46) --
  ( 45.48, 54.83) --
  ( 45.48, 54.83) --
  ( 45.48, 55.57) --
  ( 45.48, 55.57) --
  ( 45.48, 55.94) --
  ( 45.48, 55.94) --
  ( 45.48, 56.31) --
  ( 45.48, 56.31) --
  ( 45.48, 57.04) --
  ( 45.48, 57.04) --
  ( 45.48, 57.78) --
  ( 45.48, 57.78) --
  ( 45.48, 58.52) --
  ( 45.48, 58.52) --
  ( 45.48, 59.25) --
  ( 45.48, 59.25) --
  ( 45.48, 59.99) --
  ( 45.48, 59.99) --
  ( 45.48, 60.36) --
  ( 45.48, 60.36) --
  ( 45.48, 60.73) --
  ( 45.48, 60.73) --
  ( 45.48, 61.09) --
  ( 45.48, 61.09) --
  ( 45.48, 61.46) --
  ( 45.48, 61.46) --
  ( 45.48, 61.83) --
  ( 45.48, 61.83) --
  ( 45.48, 62.20) --
  ( 45.48, 62.20) --
  ( 45.48, 62.94) --
  ( 45.48, 62.94) --
  ( 45.48, 63.30) --
  ( 45.48, 63.30) --
  ( 45.48, 63.67) --
  ( 45.48, 63.67) --
  ( 45.48, 64.41) --
  ( 45.48, 64.41) --
  ( 45.48, 64.78) --
  ( 45.48, 64.78) --
  ( 45.48, 65.51) --
  ( 45.48, 65.51) --
  ( 45.48, 65.88) --
  ( 45.48, 65.88) --
  ( 45.48, 66.25) --
  ( 45.48, 66.25) --
  ( 45.48, 66.62) --
  ( 45.48, 66.62) --
  ( 45.48, 66.99) --
  ( 45.49, 66.99) --
  ( 45.49, 67.36) --
  ( 45.49, 67.36) --
  ( 45.49, 67.72) --
  ( 45.49, 67.72) --
  ( 45.49, 68.09) --
  ( 45.49, 68.09) --
  ( 45.49, 68.46) --
  ( 45.49, 68.46) --
  ( 45.49, 68.83) --
  ( 45.49, 68.83) --
  ( 45.49, 69.20) --
  ( 45.49, 69.20) --
  ( 45.49, 69.57) --
  ( 45.49, 69.57) --
  ( 45.49, 69.94) --
  ( 45.49, 69.94) --
  ( 45.49, 70.30) --
  ( 45.49, 70.30) --
  ( 45.49, 70.67) --
  ( 45.49, 70.67) --
  ( 45.49, 71.04) --
  ( 45.49, 71.04) --
  ( 45.49, 71.41) --
  ( 45.49, 71.41) --
  ( 45.49, 71.78) --
  ( 45.49, 71.78) --
  ( 45.49, 72.15) --
  ( 45.49, 72.15) --
  ( 45.49, 72.51) --
  ( 45.49, 72.51) --
  ( 45.49, 72.88) --
  ( 45.49, 72.88) --
  ( 45.49, 73.25) --
  ( 45.49, 73.25) --
  ( 45.49, 73.62) --
  ( 45.49, 73.62) --
  ( 45.49, 73.99) --
  ( 45.50, 73.99) --
  ( 45.50, 74.36) --
  ( 45.50, 74.36) --
  ( 45.50, 74.72) --
  ( 45.50, 74.72) --
  ( 45.50, 75.09) --
  ( 45.50, 75.09) --
  ( 45.50, 75.83) --
  ( 45.50, 75.83) --
  ( 45.50, 76.20) --
  ( 45.50, 76.20) --
  ( 45.50, 76.57) --
  ( 45.50, 76.57) --
  ( 45.50, 77.30) --
  ( 45.50, 77.30) --
  ( 45.50, 77.67) --
  ( 45.50, 77.67) --
  ( 45.50, 78.04) --
  ( 45.50, 78.04) --
  ( 45.50, 78.41) --
  ( 45.50, 78.41) --
  ( 45.50, 78.78) --
  ( 45.50, 78.78) --
  ( 45.50, 79.14) --
  ( 45.51, 79.14) --
  ( 45.51, 79.51) --
  ( 45.51, 79.51) --
  ( 45.51, 79.88) --
  ( 45.51, 79.88) --
  ( 45.51, 80.25) --
  ( 45.51, 80.25) --
  ( 45.51, 80.62) --
  ( 45.52, 80.62) --
  ( 45.52, 80.99) --
  ( 45.52, 80.99) --
  ( 45.52, 81.35) --
  ( 45.52, 81.35) --
  ( 45.52, 81.72) --
  ( 45.52, 81.72) --
  ( 45.52, 82.09) --
  ( 45.52, 82.09) --
  ( 45.52, 82.46) --
  ( 45.52, 82.46) --
  ( 45.52, 82.83) --
  ( 45.52, 82.83) --
  ( 45.52, 83.20) --
  ( 45.52, 83.20) --
  ( 45.52, 83.56) --
  ( 45.52, 83.56) --
  ( 45.52, 83.93) --
  ( 45.52, 83.93) --
  ( 45.52, 84.30) --
  ( 45.53, 84.30) --
  ( 45.53, 84.67) --
  ( 45.53, 84.67) --
  ( 45.53, 85.04) --
  ( 45.53, 85.04) --
  ( 45.53, 85.41) --
  ( 45.53, 85.41) --
  ( 45.53, 85.77) --
  ( 45.53, 85.77) --
  ( 45.53, 86.14) --
  ( 45.53, 86.14) --
  ( 45.53, 86.51) --
  ( 45.54, 86.51) --
  ( 45.54, 86.88) --
  ( 45.54, 86.88) --
  ( 45.54, 87.25) --
  ( 45.54, 87.25) --
  ( 45.54, 87.62) --
  ( 45.55, 87.62) --
  ( 45.55, 87.99) --
  ( 45.55, 87.99) --
  ( 45.55, 88.35) --
  ( 45.56, 88.35) --
  ( 45.56, 88.72) --
  ( 45.57, 88.72) --
  ( 45.57, 89.09) --
  ( 45.57, 89.09) --
  ( 45.57, 89.46) --
  ( 45.58, 89.46) --
  ( 45.58, 89.83) --
  ( 45.59, 89.83) --
  ( 45.59, 90.20) --
  ( 45.59, 90.20) --
  ( 45.59, 90.56) --
  ( 45.59, 90.56) --
  ( 45.59, 90.93) --
  ( 45.61, 90.93) --
  ( 45.61, 91.30) --
  ( 45.62, 91.30) --
  ( 45.62, 91.67) --
  ( 45.64, 91.67) --
  ( 45.64, 92.04) --
  ( 45.64, 92.04) --
  ( 45.64, 92.41) --
  ( 45.65, 92.41) --
  ( 45.65, 92.77) --
  ( 45.66, 92.77) --
  ( 45.66, 93.51) --
  ( 45.67, 93.51) --
  ( 45.67, 93.88) --
  ( 45.67, 93.88) --
  ( 45.67, 94.25) --
  ( 45.67, 94.25) --
  ( 45.67, 94.62) --
  ( 45.70, 94.62) --
  ( 45.70, 94.98) --
  ( 45.70, 94.98) --
  ( 45.70, 95.35) --
  ( 45.70, 95.35) --
  ( 45.70, 95.72) --
  ( 45.73, 95.72) --
  ( 45.73, 96.09) --
  ( 45.74, 96.09) --
  ( 45.74, 96.46) --
  ( 45.75, 96.46) --
  ( 45.75, 96.83) --
  ( 45.76, 96.83) --
  ( 45.76, 97.19) --
  ( 45.77, 97.19) --
  ( 45.77, 97.56) --
  ( 45.79, 97.56) --
  ( 45.79, 97.93) --
  ( 45.80, 97.93) --
  ( 45.80, 98.30) --
  ( 45.81, 98.30) --
  ( 45.81, 98.67) --
  ( 45.81, 98.67) --
  ( 45.81, 99.04) --
  ( 45.83, 99.04) --
  ( 45.83, 99.40) --
  ( 45.85, 99.40) --
  ( 45.85, 99.77) --
  ( 45.85, 99.77) --
  ( 45.85,100.14) --
  ( 45.85,100.14) --
  ( 45.85,100.51) --
  ( 45.86,100.51) --
  ( 45.86,100.88) --
  ( 45.86,100.88) --
  ( 45.86,101.25) --
  ( 45.87,101.25) --
  ( 45.87,101.61) --
  ( 45.88,101.61) --
  ( 45.88,101.98) --
  ( 45.88,101.98) --
  ( 45.88,102.35) --
  ( 45.93,102.35) --
  ( 45.93,102.72) --
  ( 45.96,102.72) --
  ( 45.96,103.09) --
  ( 45.99,103.09) --
  ( 45.99,103.46) --
  ( 46.02,103.46) --
  ( 46.02,103.82) --
  ( 46.04,103.82) --
  ( 46.04,104.19) --
  ( 46.04,104.19) --
  ( 46.04,104.56) --
  ( 46.07,104.56) --
  ( 46.07,104.93) --
  ( 46.08,104.93) --
  ( 46.08,105.30) --
  ( 46.09,105.30) --
  ( 46.09,105.67) --
  ( 46.12,105.67) --
  ( 46.12,106.04) --
  ( 46.13,106.04) --
  ( 46.13,106.40) --
  ( 46.14,106.40) --
  ( 46.14,106.77) --
  ( 46.15,106.77) --
  ( 46.15,107.14) --
  ( 46.15,107.14) --
  ( 46.15,107.51) --
  ( 46.17,107.51) --
  ( 46.17,107.88) --
  ( 46.20,107.88) --
  ( 46.20,108.25) --
  ( 46.22,108.25) --
  ( 46.22,108.61) --
  ( 46.23,108.61) --
  ( 46.23,108.98) --
  ( 46.27,108.98) --
  ( 46.27,109.35) --
  ( 46.27,109.35) --
  ( 46.27,109.72) --
  ( 46.28,109.72) --
  ( 46.28,110.09) --
  ( 46.34,110.09) --
  ( 46.34,110.46) --
  ( 46.35,110.46) --
  ( 46.35,110.82) --
  ( 46.35,110.82) --
  ( 46.35,111.19) --
  ( 46.36,111.19) --
  ( 46.36,111.56) --
  ( 46.37,111.56) --
  ( 46.37,111.93) --
  ( 46.40,111.93) --
  ( 46.40,112.30) --
  ( 46.44,112.30) --
  ( 46.44,112.67) --
  ( 46.45,112.67) --
  ( 46.45,113.03) --
  ( 46.51,113.03) --
  ( 46.51,113.40) --
  ( 46.54,113.40) --
  ( 46.54,113.77) --
  ( 46.57,113.77) --
  ( 46.57,114.14) --
  ( 46.59,114.14) --
  ( 46.59,114.51) --
  ( 46.60,114.51) --
  ( 46.60,114.88) --
  ( 46.62,114.88) --
  ( 46.62,115.24) --
  ( 46.64,115.24) --
  ( 46.64,115.61) --
  ( 46.73,115.61) --
  ( 46.73,115.98) --
  ( 46.80,115.98) --
  ( 46.80,116.35) --
  ( 46.81,116.35) --
  ( 46.81,116.72) --
  ( 46.81,116.72) --
  ( 46.81,117.09) --
  ( 46.85,117.09) --
  ( 46.85,117.45) --
  ( 46.93,117.45) --
  ( 46.93,117.82) --
  ( 46.99,117.82) --
  ( 46.99,118.19) --
  ( 47.06,118.19) --
  ( 47.06,118.56) --
  ( 47.06,118.56) --
  ( 47.06,118.93) --
  ( 47.07,118.93) --
  ( 47.07,119.30) --
  ( 47.13,119.30) --
  ( 47.13,119.66) --
  ( 47.29,119.66) --
  ( 47.29,120.03) --
  ( 47.35,120.03) --
  ( 47.35,120.40) --
  ( 47.39,120.40) --
  ( 47.39,120.77) --
  ( 47.47,120.77) --
  ( 47.47,121.14) --
  ( 47.48,121.14) --
  ( 47.48,121.51) --
  ( 47.53,121.51) --
  ( 47.53,121.87) --
  ( 47.60,121.87) --
  ( 47.60,122.24) --
  ( 47.61,122.24) --
  ( 47.61,122.61) --
  ( 47.68,122.61) --
  ( 47.68,122.98) --
  ( 47.69,122.98) --
  ( 47.69,123.35) --
  ( 47.72,123.35) --
  ( 47.72,123.72) --
  ( 47.76,123.72) --
  ( 47.76,124.09) --
  ( 47.79,124.09) --
  ( 47.79,124.45) --
  ( 47.80,124.45) --
  ( 47.80,124.82) --
  ( 47.84,124.82) --
  ( 47.84,125.19) --
  ( 47.86,125.19) --
  ( 47.86,125.56) --
  ( 47.91,125.56) --
  ( 47.91,125.93) --
  ( 48.00,125.93) --
  ( 48.00,126.30) --
  ( 48.00,126.30) --
  ( 48.00,126.66) --
  ( 48.04,126.66) --
  ( 48.04,127.03) --
  ( 48.16,127.03) --
  ( 48.16,127.40) --
  ( 48.19,127.40) --
  ( 48.19,127.77) --
  ( 48.30,127.77) --
  ( 48.30,128.14) --
  ( 48.34,128.14) --
  ( 48.34,128.51) --
  ( 48.48,128.51) --
  ( 48.48,128.87) --
  ( 48.50,128.87) --
  ( 48.50,129.24) --
  ( 48.80,129.24) --
  ( 48.80,129.61) --
  ( 48.86,129.61) --
  ( 48.86,129.98) --
  ( 49.26,129.98) --
  ( 49.26,130.35) --
  ( 49.35,130.35) --
  ( 49.35,130.72) --
  ( 49.63,130.72) --
  ( 49.63,131.08) --
  ( 49.64,131.08) --
  ( 49.64,131.45) --
  ( 49.66,131.45) --
  ( 49.66,131.82) --
  ( 49.74,131.82) --
  ( 49.74,132.19) --
  ( 49.84,132.19) --
  ( 49.84,132.56) --
  ( 49.93,132.56) --
  ( 49.93,132.93) --
  ( 50.15,132.93) --
  ( 50.15,133.29) --
  ( 50.22,133.29) --
  ( 50.22,133.66) --
  ( 50.27,133.66) --
  ( 50.27,134.03) --
  ( 50.34,134.03) --
  ( 50.34,134.40) --
  ( 50.38,134.40) --
  ( 50.38,134.77) --
  ( 50.46,134.77) --
  ( 50.46,135.14) --
  ( 50.48,135.14) --
  ( 50.48,135.50) --
  ( 50.48,135.50) --
  ( 50.48,135.87) --
  ( 50.49,135.87) --
  ( 50.49,136.24) --
  ( 50.54,136.24) --
  ( 50.54,136.61) --
  ( 50.65,136.61) --
  ( 50.65,136.98) --
  ( 50.91,136.98) --
  ( 50.91,137.35) --
  ( 50.95,137.35) --
  ( 50.95,137.71) --
  ( 51.10,137.71) --
  ( 51.10,138.08) --
  ( 51.23,138.08) --
  ( 51.23,138.45) --
  ( 51.63,138.45) --
  ( 51.63,138.82) --
  ( 51.89,138.82) --
  ( 51.89,139.19) --
  ( 52.37,139.19) --
  ( 52.37,139.56) --
  ( 52.55,139.56) --
  ( 52.55,139.92) --
  ( 52.79,139.92) --
  ( 52.79,140.29) --
  ( 52.82,140.29) --
  ( 52.82,140.66) --
  ( 53.40,140.66) --
  ( 53.40,141.03) --
  ( 53.55,141.03) --
  ( 53.55,141.40) --
  ( 53.67,141.40) --
  ( 53.67,141.77) --
  ( 54.03,141.77) --
  ( 54.03,142.14) --
  ( 55.23,142.14) --
  ( 55.23,142.50) --
  ( 55.52,142.50) --
  ( 55.52,142.87) --
  ( 55.57,142.87) --
  ( 55.57,143.24) --
  ( 55.68,143.24) --
  ( 55.68,143.61) --
  ( 55.86,143.61) --
  ( 55.86,143.98) --
  ( 55.99,143.98) --
  ( 55.99,144.35) --
  ( 56.00,144.35) --
  ( 56.00,144.71) --
  ( 56.18,144.71) --
  ( 56.18,145.08) --
  ( 56.87,145.08) --
  ( 56.87,145.45) --
  ( 56.98,145.45) --
  ( 56.98,145.82) --
  ( 57.19,145.82) --
  ( 57.19,146.19) --
  ( 57.22,146.19) --
  ( 57.22,146.56) --
  ( 57.38,146.56) --
  ( 57.38,146.92) --
  ( 57.74,146.92) --
  ( 57.74,147.29) --
  ( 57.76,147.29) --
  ( 57.76,147.66) --
  ( 57.79,147.66) --
  ( 57.79,148.03) --
  ( 57.88,148.03) --
  ( 57.88,148.40) --
  ( 58.61,148.40) --
  ( 58.61,148.77) --
  ( 58.62,148.77) --
  ( 58.62,149.13) --
  ( 58.63,149.13) --
  ( 58.63,149.50) --
  ( 58.69,149.50) --
  ( 58.69,149.87) --
  ( 58.84,149.87) --
  ( 58.84,150.24) --
  ( 59.09,150.24) --
  ( 59.09,150.61) --
  ( 60.58,150.61) --
  ( 60.58,150.98) --
  ( 61.25,150.98) --
  ( 61.25,151.34) --
  ( 61.29,151.34) --
  ( 61.29,151.71) --
  ( 62.02,151.71) --
  ( 62.02,152.08) --
  ( 64.17,152.08) --
  ( 64.17,152.45) --
  ( 64.73,152.45) --
  ( 64.73,152.82) --
  ( 64.91,152.82) --
  ( 64.91,153.19) --
  ( 65.93,153.19) --
  ( 65.93,153.55) --
  ( 66.63,153.55) --
  ( 66.63,153.92) --
  ( 67.07,153.92) --
  ( 67.07,154.29) --
  ( 68.86,154.29) --
  ( 68.86,154.66) --
  ( 69.23,154.66) --
  ( 69.23,155.03) --
  ( 69.38,155.03) --
  ( 69.38,155.40) --
  ( 70.05,155.40) --
  ( 70.05,155.76) --
  ( 71.28,155.76) --
  ( 71.28,156.13) --
  ( 71.48,156.13) --
  ( 71.48,156.50) --
  ( 72.49,156.50) --
  ( 72.49,156.87) --
  ( 73.23,156.87) --
  ( 73.23,157.24) --
  ( 74.40,157.24) --
  ( 74.40,157.61) --
  ( 75.39,157.61) --
  ( 75.39,157.97) --
  ( 76.89,157.97) --
  ( 76.89,158.34) --
  ( 81.65,158.34) --
  ( 81.65,158.71) --
  ( 82.57,158.71) --
  ( 82.57,159.08) --
  ( 83.63,159.08) --
  ( 83.63,159.45) --
  ( 84.44,159.45) --
  ( 84.44,159.82) --
  ( 84.54,159.82) --
  ( 84.54,160.19) --
  ( 85.69,160.19) --
  ( 85.69,160.55) --
  ( 87.08,160.55) --
  ( 87.08,160.92) --
  ( 88.88,160.92) --
  ( 88.88,161.29) --
  ( 92.46,161.29) --
  ( 92.46,161.66) --
  ( 93.68,161.66) --
  ( 93.68,162.03) --
  ( 95.62,162.03) --
  ( 95.62,162.40) --
  ( 96.14,162.40) --
  ( 96.14,162.76) --
  ( 98.28,162.76) --
  ( 98.28,163.13) --
  ( 98.54,163.13) --
  ( 98.54,163.50) --
  (105.30,163.50) --
  (105.30,163.87) --
  (109.93,163.87) --
  (109.93,164.24) --
  (111.73,164.24) --
  (111.73,164.61) --
  (111.77,164.61) --
  (111.77,164.97) --
  (112.08,164.97) --
  (112.08,165.34) --
  (112.28,165.34) --
  (112.28,165.71) --
  (116.28,165.71) --
  (116.28,166.08) --
  (119.53,166.08) --
  (119.53,166.45) --
  (123.36,166.45) --
  (123.36,166.82) --
  (136.76,166.82) --
  (136.76,167.18) --
  (139.86,167.18) --
  (139.86,167.55) --
  (140.22,167.55) --
  (140.22,167.92) --
  (156.60,167.92) --
  (156.60,168.29) --
  (185.59,168.29) --
  (185.59,168.66) --
  (203.89,168.66) --
  (203.89,169.03) --
  (211.81,169.03) --
  (211.81,169.03);

\path[draw=drawColor,line width= 0.6pt,dash pattern=on 1pt off 3pt ,line join=round] ( 95.62, 29.40) -- ( 95.62,175.68);

\node[text=drawColor,anchor=base,inner sep=0pt, outer sep=0pt, scale=  1.10] at (107.47, 32.24) {95\% quantile};
\definecolor{drawColor}{gray}{0.20}

\path[draw=drawColor,line width= 0.6pt,line join=round,line cap=round] ( 37.55, 29.40) rectangle (211.81,175.68);
\end{scope}
\begin{scope}
\path[clip] (  0.00,  0.00) rectangle (216.81,180.67);
\definecolor{drawColor}{gray}{0.30}

\node[text=drawColor,anchor=base east,inner sep=0pt, outer sep=0pt, scale=  0.80] at ( 33.05, 33.29) {0.00};

\node[text=drawColor,anchor=base east,inner sep=0pt, outer sep=0pt, scale=  0.80] at ( 33.05, 66.54) {0.25};

\node[text=drawColor,anchor=base east,inner sep=0pt, outer sep=0pt, scale=  0.80] at ( 33.05, 99.78) {0.50};

\node[text=drawColor,anchor=base east,inner sep=0pt, outer sep=0pt, scale=  0.80] at ( 33.05,133.03) {0.75};

\node[text=drawColor,anchor=base east,inner sep=0pt, outer sep=0pt, scale=  0.80] at ( 33.05,166.27) {1.00};
\end{scope}
\begin{scope}
\path[clip] (  0.00,  0.00) rectangle (216.81,180.67);
\definecolor{drawColor}{gray}{0.20}

\path[draw=drawColor,line width= 0.6pt,line join=round] ( 35.05, 36.05) --
  ( 37.55, 36.05);

\path[draw=drawColor,line width= 0.6pt,line join=round] ( 35.05, 69.29) --
  ( 37.55, 69.29);

\path[draw=drawColor,line width= 0.6pt,line join=round] ( 35.05,102.54) --
  ( 37.55,102.54);

\path[draw=drawColor,line width= 0.6pt,line join=round] ( 35.05,135.78) --
  ( 37.55,135.78);

\path[draw=drawColor,line width= 0.6pt,line join=round] ( 35.05,169.03) --
  ( 37.55,169.03);
\end{scope}
\begin{scope}
\path[clip] (  0.00,  0.00) rectangle (216.81,180.67);
\definecolor{drawColor}{gray}{0.20}

\path[draw=drawColor,line width= 0.6pt,line join=round] ( 45.46, 26.90) --
  ( 45.46, 29.40);

\path[draw=drawColor,line width= 0.6pt,line join=round] ( 91.04, 26.90) --
  ( 91.04, 29.40);

\path[draw=drawColor,line width= 0.6pt,line join=round] (136.62, 26.90) --
  (136.62, 29.40);

\path[draw=drawColor,line width= 0.6pt,line join=round] (182.20, 26.90) --
  (182.20, 29.40);
\end{scope}
\begin{scope}
\path[clip] (  0.00,  0.00) rectangle (216.81,180.67);
\definecolor{drawColor}{gray}{0.30}

\node[text=drawColor,anchor=base,inner sep=0pt, outer sep=0pt, scale=  0.80] at ( 45.46, 19.39) {0};

\node[text=drawColor,anchor=base,inner sep=0pt, outer sep=0pt, scale=  0.80] at ( 91.04, 19.39) {5};

\node[text=drawColor,anchor=base,inner sep=0pt, outer sep=0pt, scale=  0.80] at (136.62, 19.39) {10};

\node[text=drawColor,anchor=base,inner sep=0pt, outer sep=0pt, scale=  0.80] at (182.20, 19.39) {15};
\end{scope}
\begin{scope}
\path[clip] (  0.00,  0.00) rectangle (216.81,180.67);
\definecolor{drawColor}{RGB}{0,0,0}

\node[text=drawColor,anchor=base,inner sep=0pt, outer sep=0pt, scale=  1.00] at (124.68,  7.50) {Duration [day]};
\end{scope}
\begin{scope}
\path[clip] (  0.00,  0.00) rectangle (216.81,180.67);
\definecolor{drawColor}{RGB}{0,0,0}

\node[text=drawColor,rotate= 90.00,anchor=base,inner sep=0pt, outer sep=0pt, scale=  1.00] at ( 11.89,102.54) {Cumulative Frequency};
\end{scope}
\end{tikzpicture}

  \caption{Cumulative distribution of durations in the trace. \SI{95}{\percent} of the considered durations where shorter than \SI{5.5}{\day}.}
  \label{fig:cumulativeDurations}
\end{figure}

Since recently restarted nodes exhibit low-duration connections and high churn, the question arises if the same behavior is true for long-running nodes, \ie, how connection durations are distributed.
To this end, we ran a unaltered Geth node for \SI{18.76}{\day} (450 hours) and logged the duration of every connection.
The results are depicted in \Cref{fig:cumulativeDurations}, showing the cumulative distribution of durations in the trace.
To improve the readability of the Figure, we excluded connections with a duration shorter than \SI{60}{\second}, which were \SI{90.26}{\percent} of all connections, yielding a total of 361 connections%
\footnote{
We suspect the heterogeneity of the discovery table as the major cause for the abundance of connections shorter than \SI{60}{\second}~\cite{DBLP:conf/imc/KimMMMMB18}.
Ethereum is embedded in a family of protocols (\Cref{sec:networkLayer}), all of which use the same ports and message structures.
\textcite{DBLP:conf/imc/KimMMMMB18} report that only \SI{10}{\percent} of node IDs in the discovery table correspond to peers speaking the Ethereum protocol, out of which only \SI{50}{\percent} operate on the Ethereum main network.
Hence, in the majority of situations DEVp2p tries to establish connections to peers which are not useful and immediately discards them again.
}.

It can be seen in \Cref{fig:cumulativeDurations} that the longest connection duration was \SI{60}{\day}, but the majority of connections was much shorter lived.
The quantile shows that \SI{95}{\percent} of the considered connections were shorter than \SI{5.5}{\day}; only 18 connections where longer than this duration.
Though the peer churn is lower than in a restarted node, it is still surprisingly high for which we have several conjectures:
Geth's development cycle is fast and requires frequent updates, either due to security fixes or protocol changes.
Additionally, read (write) timeouts of \SI{20}{s} (\SI{30}{s}) on the TCP-level are relatively small compared to other networks like Bitcoin.
On the UDP-level, timeouts are currently set at \SI{500}{ms}, while~\cite{DBLP:journals/corr/abs-1801-03998} report average inter-node latencies of roughly \SI{180}{ms}, with \SI{10}{\percent} of peers having a latency higher than \SI{276}{ms}.
Consequently, buckets in the discovery table experience a high level of churn, making it easy to enter even fully filled ones.

Given that most connections on the Ethereum network are rather short-lived, we conducted a proof-of-concept attack without restarting the victim.
We let the victim node run without any attack activity for 72 hours to populate the discovery table, mimicking a more realistic network state.
In our experiment the false friends attack was successful after \SI{4.765}{\day} (114.4 hours).
\footnote{In two subsequent experiments, the victim was left with only one benign connection after \SI{4.875}{\day} and \SI{9.5}{\day}, respectively (i.e., the node was only one connection away from being fully eclipsed).}
\section{Countermeasures}
\label{sec:countermeasures}

Several modifications to Geth are conceivable, some quickly realizable and some more fundamental, to counteract the false friends eclipse.
In the following we propose several countermeasures and subsequently describe the ones that where implemented in Geth.

\subsection{Proposed Countermeasures}

First, we focus on quickly realizable modifications that would immediately increase the cost of a successful false friends eclipse.
One possibility are more stringent \emph{IP subnet restrictions} in the discovery table and on the DEVp2p connection layer.
While increasing the bucket size is not promising (cf. \Cref{subsec:findnode_probability}), enforcing \emph{any} subnet restriction on the replies of FindNode-Requests would increase the number of unique IP adresses necessary for a successful attack.
Another low-invasive countermeasure is to consider all known nodes in \texttt{ReadRandomNodes}, instead of only the heads of each bucket.
Furthermore, we are only able to perform our attack without requiring a restart of the victim node because peering relationships in Ethereum are currently very short-lived.
Increasing timeouts on both the TCP- and UDP-level could decrease this volatility.

On a more fundamental level, we argue that the \emph{complexity} of Geth's current node selection logic is a major enabler for attacks such as~\cite{DBLP:journals/iacr/MarcusHG18} and our false friends attack.
New peers are chosen based on their node ID, which arguably does not make any sense if the goal of the resulting overlay is flooding identical information to all nodes (in contrast to ID-based routing).
Node IDs are, however, trivially manipulatable by adversaries to optimize the placement of adversary nodes in peer discovery tables.
The complexities of ID-based peer selection are therefore not only unnecessary, but also detrimental to security.
For a sustainable, long-term fix we strongly suggest to ignore node IDs for all aspects of peer discovery for the Ethereum protocol.
Instead, peering decisions should be weighted by more expensive-to-manipulate node characteristics, such as IP addresses or, perhaps, publicly locked Ether stake linked to individual nodes.

Decades of research on Sybil-attacks in peer-to-peer-networks~\cite{douceur2002sybil} suggest that in a completely trustless setting it is only viable to make the creation of a multitude of adversarial nodes expensive, not impossible.
The implications of this are twofold.
First, in the typical blockchain scenario one honest node is sufficient to prevent an eclipse attack.
Therefore, the probability of filling the whole peer list with adversarial nodes must be minimized by means that are robust to a potentially substantial population of Sybil nodes.
We suggest that a promising approach for achieving this is a combination of maintaining a large peer set and using a node selection process which as closely as possible resembles a uniform draw from the set of all network nodes.
For example, Geth's complex Kademlia-inspired bucket structure and discovery logic can be replaced with a single data structure holding a large number of nodes (peer candidates) from which peers are drawn uniformly at random.
Second, nodes that are profitable targets for Eclipse attacks (high-profile merchants, miners) should not rely on a purely trustless node selection logic.
Instead, these nodes should statically include known and trusted nodes into their peer list, as seems to be practice in the Bitcoin network~\cite{DBLP:conf/uss/HeilmanKZG15}.
In other words, potentially attractive targets might want to invest manual effort to \emph{choose their friends wisely}.

\subsection{Implemented Countermeasures}

In response to our responsible disclosure and subsequent discussions\footnote{We want to thank Felix Lange for his time and the fruitful discussions.} the Ethereum developers implemented several low-invasive countermeasures that have been incorporated into the 1.9.0-release of Geth.
These changes mitigate the immediate threat of false friends eclipse attacks and raise the bar for an attacker by requiring an increased number of Sybil nodes to carry out an attack.
Admittedly, we still strongly suggest to fade out the structured Kademlia-based discovery for a less structured approach, \eg, the addrman used in Bitcoin.

\subsubsection{Raising the Number of Connections to 50}

Pre Geth v1.9.0, each node established a total of 25 connections by default, 8 of which are outbound and 17 are inbound.
The number of outbound connections in Geth is defined as
\begin{align*}
  \text{\# of outbound} = \lfloor \frac{\text{max peers}}{3} \rfloor = \lfloor \frac{25}{3} \rfloor = 8.
\end{align*}

The total number of connections a Geth node establishes has been double from 25 to 50, effectively doubling the number of outbound connections as well.
Therefore, an adversary would require more resources to successfully eclipse a victim.
Furthermore, since the false friends eclipse is only successful when existing connections to honest nodes are dropped, the increased limit raises the chances of maintaining a long-lived connection to an honest node, thwarting the eclipse as long as these connections are sustained.

\subsubsection{\texttt{ReadRandomNodes} Considers All Nodes}

The function \texttt{ReadRandomNodes} now selects nodes uniformly at random from the set of all nodes in the table, instead of just the bucket heads.
An adversary would have to overtake the whole table to deterministically ensure that only adversarial nodes are returned.
However, the maximum number of nodes in the table is relatively small: 17 buckets with a maximum of 16 nodes, \ie, 272 possible nodes in total.
Assuming 18 adversarial nodes as used throughout this paper, the chances of selection an attacker-node at random are $\frac{18}{272} \approx 7\%$.
In practice, having a full table is very unlikely, due to Kademlias distance metric: the size of the potential node-set of each bucket decreases exponentially.
While the first buckets are always filled, the later ones tend to be almost empty most of the time.
\textcite{DBLP:journals/iacr/MarcusHG18} report an average table population of 168 nodes, therefore increasing the chances of randomly selecting an adversarial node to $\approx 11\%$.

We argue that while selecting peers uniformly at random is a desirable strategy with respect to robustness, the node set from which peers are drawn should be sufficiently large.

\subsubsection{Throttle Inbound Connection Attempts}

A major facilitator of our false friends eclipse is the ability to establish inbound connections from the same IP address.
All inbound connection slots could be filled during our evaluation by running just one server.
Since IP addresses are the most costly part in such an attack (in comparison to memory or computational power), increasing the necessary number of addresses to fill inbound connection slots is vital to raise the bar for an attacker.

Since Geth v1.9.0, inbound connection attempts from the same IP now have to wait \SI{30}{\second}.
While this is a first step in raising the bar for an attacker, we argue that this is not enough.
Additional subnet restrictions on inbound connections (\eg, only 2 IPs from the same /24 subnet) are an effective and low-invasive way of making an eclipse more difficult.

\subsubsection{Unchanged: The lookup-buffer}

A major component of our false friends attack is the exploitation of the lookup buffer (cf. \cref{subsec:findnodeProb}).
With one adversarial node in each bucket (which is still possible) and a sufficiently large number of pre-computed node IDs, an adversary can still ensure that the lookup-buffer is filled with adversarial nodes.
Since \texttt{ReadRandomNodes} cannot be exploited without a significant resource-investment, a compromised lookup-buffer does not pose an immediate threat.
However, we strongly suggest to enforce subnet restrictions or ignoring the lookup-buffer completely for peer-selection
\footnote{These things will potentially be addressed in future releases of Geth.}.

\section{Related Work}
\label{sec:rw}

The literature on security considerations in peer-to-peer networks in general and Kademlia-based networks in particular is vast; in the following we focus on research most closely related to our attack instead of providing a broad overview.

\subsection{Attacks and Countermeasures in Overlay Networks}

\textcite{DBLP:conf/infocom/SinghNDW06,DBLP:conf/osdi/CastroDGRW02,DBLP:conf/iptps/SitM02} survey eclipse attacks and countermeasures on peer-to-peer overlay networks.
Similar to our reasoning, \cite{DBLP:conf/infocom/SinghNDW06,DBLP:conf/osdi/CastroDGRW02} conjecture that Sybil attacks cannot be solved in a purely peer-to-peer fashion.
As a remedy, they propose a central trusted certificate authority to bind node IDs to identities which subsequently enables the implementation of countermeasures not possible in purely trustless systems.
A decentralized mitigation is proposed and thoroughly analyzed in~\cite{DBLP:conf/cns/GermanusRSS14}, which is based on purposely letting lookups diverge, so that an adversary cannot easily eclipse a target by poisoning its logical proximity.

\subsection{Attacks on Kademlia-based networks}

The security Kademlia~\cite{DBLP:conf/iptps/MaymounkovM02} and its inspired implementations have been studied extensively~\cite{DBLP:journals/ccr/SteinerEB07,DBLP:conf/icdcn/LocherMSW10,DBLP:conf/networking/KohnenLR09,DBLP:journals/scn/WangTCMKHK13}.
\textcite{DBLP:journals/ccr/SteinerEB07} explore the space of possible attacks and implications whereas subsequent works focus on optimizations of these attacks~\cite{DBLP:conf/icdcn/LocherMSW10,DBLP:journals/scn/WangTCMKHK13} and circumventing implemented countermeasures~\cite{DBLP:conf/networking/KohnenLR09}.
Most approaches require the ability to arbitrarily choose node IDs.
Similar to our false friend attack where we insert carefully selected node IDs into the victim's discovery table, \cite{DBLP:journals/scn/WangTCMKHK13} present a low-resource approach to poison routing entries in the KAD network.
Given multiple attacking nodes, the ID space is partitioned and routing entries hijacked by spoofing messages.
In Ethereum, message spoofing and arbitrary node ID choice are impossible, making our attack conceptually different, though closely related to previous attacks on the KAD network.
Most notably, \cite{DBLP:conf/icdcn/LocherMSW10} also conjecture that a purely trustless countermeasure cannot exist, due to the fundamental problem of Sybil identities~\cite{douceur2002sybil}.

\subsection{Eclipse Attacks in Blockchain Systems}

\textcite{DBLP:conf/uss/HeilmanKZG15} were the first to study eclipse attacks on peer-to-peer Blockchain-systems, in particular Bitcoin.
Despite the introduced countermeasures, eclipse attacks are still possible when exploiting BGP~\cite{DBLP:conf/sp/ApostolakiZV17}.
For Ethereum, \cite{gervais_ethereum_2016} describe an attack on the block synchronization mechanism.
When an Ethereum peer misses a block, it will start a synchronization with exactly one neighboring peer.
An adversary can leverage this behavior to indefinitely stall the synchronization or inject an adversarial chain of blocks.

As noted throughout this paper, several eclipse attacks on Ethereum are described in \cite{DBLP:journals/iacr/MarcusHG18}.
Our approach differs since we do not fill the complete table with adversarial nodes instead insert node IDs with specific properties.

\section{Conclusion}
\label{sec:conclusion}

We presented the \emph{false friends} attack, an eclipse attack applicable to current versions of Geth, the by far most popular Ethereum node software.
Our attack requires only $2$ IPs from distinct /24 subnets to be successful.
Moreover, and in contrast to previous attacks, it can be successfully mounted without assuming that the victim node reboots at some point.
Empirical measurements of our attack in the live Ethereum mainnet indicate that even without a restart of the victim node, a false friends eclipse can be completed in a matter of days.
Our discovery is even more striking when considering that countermeasures against similar attacks were only recently introduced to the Geth codebase.
We argue that the ongoing vulnerability of Geth is at least partly due to a fundamentally unsuited node discovery approach.
While we propose both short- and long-term countermeasures to the false friends attack, existing literature hints that in a completely trustless setting, eclipse attacks can only be made expensive, not impossible.
Potentially attractive targets might wish to invest manual effort towards choosing their friends wisely.

\section*{Acknowledgements}

We would like to thank to Holger Döbler, Roman Naumann, Elias Rohrer, Samuel Brack, Till Neudecker and Florian Tschorsch for the highly valuable discussions on the topic.
Additionally, we want to thank Felix Lange for his efforts and openness to our suggestions during the fruitful discussions.

\printbibliography
\end{document}